\documentstyle[epsfig]{mn}

\newcommand{\iras} {{\it IRAS \/}}

\newcommand{\be}{\begin{equation}}
\newcommand{\ee}{\end{equation}}

\newcommand{\bea}{\begin{eqnarray}}
\newcommand{\eea}{\end{eqnarray}}

\newcommand{\bc}{\begin{center}}
\newcommand{\ec}{\end{center}}

\newcommand{\lu}{\,h^{-1}{\rm Mpc}}

\renewcommand{\vec}[1]{ {\bmath #1} } 

\newcommand{\dd}{{\rm d}}

\title{The Topology of the IRAS Point Source Catalogue Redshift Survey}

\author[A. Canavezes et al.]
{A. Canavezes,$^1$ V. Springel,$^2$ S. J. Oliver,$^1$
M. Rowan-Robinson,$^1$ O. Keeble,$^1$\cr
S. D. M. White,$^2$ W. Saunders,$^3$ G. Efstathiou,$^4$ C. Frenk,$^5$
R. G. McMahon,$^4$\cr
S. Maddox,$^4$ W. Sutherland,$^6$ and H. Tadros$^7$ \\
$^1$Imperial College of Science Technology and Medicine, Blackett
Laboratory, Prince Consort Road. London SW7 2BZ, UK \\
$^2$Max-Planck-Institut f\"{u}r Astrophysik, 
Karl-Schwarzschild-Stra\ss{}e 1, 
85740 Garching bei M\"{u}nchen, Germany\\
$^3$Institute for Astronomy, University of Edinburgh, Blackford Hill, Edinburgh EH9 3JS, UK\\
$^4$Institute of Astronomy, University of Cambridge, Madingley Road, Cambridge CB3 OHA, UK\\
$^5$Department of Physics, University of Durham, South Road, Durham, DH1 3LE, UK\\
$^6$Department of Physics, University of Oxford, Keeble Road, Oxford OX1 3RH, UK\\
$^7$Department of Physics, University of Sussex, Falmer, Brighton BN1 9QH, UK}

\begin{document}
\maketitle 

\begin{abstract}

	We investigate the topology of the new Point Source Catalogue 
Redshift Survey (PSCz) of {\it IRAS} 
galaxies by means of the genus statistic. The survey maps the local
Universe with approximately 15000 galaxies over 
84.1 per cent of the sky and provides
an unprecedented number of resolution elements for the topological
analysis. For comparison with the PSCz data we also examine the genus of  
large N-body simulations of four variants of the cold dark matter
cosmogony.
The simulations are part of the Virgo project to simulate 
the formation of structure in the Universe.
We assume that the statistical properties of the 
galaxy distribution can be identified with those of 
the dark matter particles in the simulations. 
We extend the standard genus analysis by 
examining the influence of sampling noise on the genus curve and 
introducing a statistic able to quantify the amount 
of phase correlation present in the density field, 
the {\it amplitude drop} of the genus compared to
a Gaussian field with identical power spectrum. 
The results for PSCz are consistent with the hypothesis of 
random phase initial conditions. In particular, no strong 
phase correlation is detected on scales 
ranging from $10\lu$ to $32\lu$, whereas there is a positive detection
of phase correlation at smaller scales.
Among the simulations,
phase correlations are 
detected in all models at small scales, albeit with different strengths.
When scaled to a common normalization, 
the amplitude drop primarily depends on the shape of the power
spectrum. We find that 
the constant bias standard CDM model 
can be ruled out
at high significance 
because the shape of its 
power spectrum is not consistent with PSCz. The other CDM models with
more large-scale power all fit
the PSCz data almost equally well, with a slight preference for 
a high density $\tau$CDM model.

\end{abstract}


\section{Introduction}

        \label{intro}
	
All-sky redshift surveys of galaxies selected from the {\it IRAS} catalogues
have had a dramatic impact on our understanding of the large scale
structure and dynamics of the local Universe.  The QDOT survey
\cite{La95} provided early evidence that the
large-scale clustering of galaxies was incompatible with the standard
cold dark matter model \cite{Ef90,Sa91}. 
Comparison of our Local Group motion or the local
velocity field as determined from peculiar velocity studies, with the
velocity field inferred from the {\it IRAS} galaxy density, has provided
determinations of $\Omega^{0.6}/b$ on very large scales
\cite{Ro90,Ka91,De93,Nu94,St92b}.  These
dynamical studies were in general compatible with $\Omega=1$
cosmogonies with only small deviations of the bias parameter from
$b=1$.  This suggested that {\it IRAS} galaxies are reasonably faithful
tracers of the mass field, as compared to e.g.\ optically selected
galaxies.  The main advantage of {\it IRAS} galaxy redshift surveys is,
however, the uniform all-sky coverage as a result of the {\it IRAS} survey
planning and the reduced Galactic obscuration in the far infrared.
{\it IRAS} galaxy surveys were also able to probe cosmologically significant
volumes efficiently due to their broad selection function.

The two most significant {\it IRAS} redshift surveys to date employed two
different strategies.  The QDOT survey consists of 2387 galaxies
randomly selected at a rate of 1 in 6 to the full depth of the {\it IRAS}
Point Source Catalog (PSC), 0.6$\,$Jy, thus exploring the maximum volume
at the expense of sampling density. The 1.2-Jy survey \cite{Fi95}, an 
extension of the 2-Jy survey \cite{St92a},
comprises 5321 galaxies fully sampled from the PSC but to a shallower
depth.  

The sparse sampling strategy of QDOT was ideal for low order
statistics such as the galaxy power spectrum whereas the 1.2-Jy survey was
more useful in measurements of higher order statistics which require
higher sampling densities \cite{Fi94,Nu95,Bo93}. 
Some important studies such as
the inferred local velocity field, the convergence of the dipole and
the topology greatly benefit from both high sampling densities and
cosmologically significant volumes. 
For these reasons we have constructed
the PSCz \cite{Sa94}, a complete redshift catalogue of 15000 galaxies to the
full depth of the PSC (0.6$\,$Jy).

The construction of the parent sample for the PSCz was similar to
that of the QMW {\it IRAS} Galaxy Catalogue \cite{Ro90}
with substantial efforts made to improve the sky coverage, uniformity
and completeness. Details are provided in Saunders et
al.\ \shortcite{Sa97}.  
Optical
counterparts were identified using digitized plate material from APM
and COSMOS. Galactic sources were excluded on the basis of {\it IRAS}
colours and literature searches.  Existing surveys and literature
material provided around two thirds of the redshifts, allowing the
full redshift catalogue to be constructed with realistic amounts of
telescope time. We have now measured around 5000 new redshifts to
provide a final catalogue with redshift completeness of 98 per cent;
the observing  programme and data reduction are described
in Keeble et~al. \shortcite{Ke97}.

There has been extensive 
discussion in the literature on the ability 
of topological studies to discriminate between 
different models of structure formation; not only 
between inflationary models 
(with random-phase initial conditions) 
and non-Gaussian models of structure formation, 
like cosmic strings or textures, 
but also between models with different 
dark matter content \cite{Ker97,Ma96c,Co96}.
From these previous studies, it is clear that the genus statistic is a
robust discriminator as long as low-noise data are used. 
Our use of very large simulations of 
CDM variants in conjunction with the PSCz survey is 
ideal to reveal the power of this 
statistic in discriminating between different models 
of structure formation. 

Topological studies have also been aimed at drawing conclusions on the
shape of the power spectrum. Moore et~al.\ \shortcite{Mo92} concluded
that 
the amplitude of the genus curves on large scales was inconsistent
with 
the predictions of a constant bias standard CDM model and that the
power 
spectrum was best fit with a 
power law with index $n=-1$. Vogeley et al. \shortcite{Vo94} 
showed that the values for the amplitude of the genus curves were 
again inconsistent with a constant bias standard CDM model but
consistent 
with an open CDM model, 
and Protogeros \& Weinberg \shortcite{Pr97} concluded 
that the topology of the 1.2-Jy redshift survey was best fit by a
power 
spectrum with index $n=-1$. 
However, the quantitative significance of these results was not up to 
what one would wish, given the limitations of the data they used.

This paper forms part of a sequence describing the first results of the
PSCz survey:  Saunders et~al.\ \shortcite{Sa97} describe the cosmography and
analyse the counts-in-cells statistics;  Rowan-Robinson et~al.\ \shortcite{Ro97}
examine the convergence of the density and velocity dipole;
Sutherland et~al.\ \shortcite{Su97} 
discuss the power spectrum; Tadros et~al.\ \shortcite{Ta97} 
perform a spherical harmonic analysis of redshift space distortions, while
Keeble et~al.\ \shortcite{Ke97}  compare the radial and transverse components 
of the correlation function. Here we analyse the topology of the
PSCz density field: 
Section~\ref{topo} discusses the topological
statistics applied
while Section~\ref{denmaps} describes the construction 
of the PSCz density map.
We present 
results for PSCz 
in Section~\ref{results} and give an 
account of the theoretical models used to compare with the
data in Section~\ref{nbody}. Finally, 
we present our conclusions in Section~\ref{conclusion}.

\section{Topological Methods}

        \label{topo}
	
\subsection{Genus statistics}
 
The statistical tools currently used in order 
to confront theories of structure formation with observational data 
go well beyond the simple two-point correlation function. 
The search for such `higher-order' information is based 
on the realization that the two-point correlation function 
(or equivalently the power spectrum) exhausts the 
statistical content of a system only when this has a Gaussian nature. 
It is clear, however, that non-linear gravitational clustering will
inevitably introduce
non-Gaussian features, even allowing for random-phase initial
conditions.

In order to measure departures from Gaussianity, 
Gott, Melott \& Dickinson \shortcite{Go86} suggested
studying the topology of isodensity surfaces and 
quantifying it via the {\it genus}. Subsequently the genus statistic has
been applied to a number of redshift surveys
\cite{Go89,Mo92,Pa92,Vo94,Rh94,Pr97} and has been subject to several
theoretical investigations \cite{Ha86,We87,Me88,Pa91,Be92a,Go96,Ma94,Ma96a,Ma96b,Ma96c}.
The genus has also been considered in the more general
framework of Minkowski functionals \cite{Me94}.

Given an isodensity surface $S$, 
the genus $G$ of that surface can be defined as
\be
 G(S) = \: \mbox{\# of holes} \: -  \: \mbox{\# of isolated
 regions} \: +1 ,
\ee
i.e. a spherical surface has genus 0; 
a torus has genus 1, 
whereas a distribution of $N$ 
disjoint spherical surfaces gives rise to ${\it G}=-(N-1)$.

A more formal definition of the genus 
can be given by means of the Gauss-Bonnet theorem, 
which relates the curvature of the surface to the genus. 
Let $k=(a_{1}a_{2})^{-1}$ 
be the local Gaussian curvature of a two 
dimensional surface $S$, i.e. $a_1$ and $a_2$ are the
two principal radii of curvature.
Then, the Gauss-Bonnet theorem states that
\be
G(S)=-\frac{1}{4\pi} \int_{S} k \, {\rm d}A +1 .
\ee

Defining $S$
as the boundary surface between two regions above and below a 
threshold density $\rho_{\rm t}$ 
we can calculate $G(S)$ 
and plot it as a function of that threshold, 
thus obtaining the {\it genus curve}\, of a given density field.

What do we expect the genus curve to look like? 
If a high threshold is selected, 
only a few very dense and isolated regions will be above this
density value and the genus is negative. 
If a low threshold density is chosen, 
only a few isolated voids are identified and, again, 
the genus is negative. 
On the other hand, for a threshold around the mean density value, 
one expects, in general, that the isodensity surfaces will have a
multiply 
connected structure that resembles a sponge, 
with a resulting positive genus.

The simplest case occurs, of course, for a Gaussian random field. 
In this case, the statistical properties of the density field are 
completely described by the power spectrum and so the genus will not 
provide any additional information. Moreover, underdense or overdense 
regions are statistically indistinguishable. This means the genus curve is symmetric
about the mean, which is characteristic of the so-called
`sponge-like' 
topology. Parameterizing the threshold density by the
number $\nu \equiv (\rho -\rho_{\rm mean})/\sigma_{\rho}$
of standard deviations from 
the mean density, 
and introducing the genus per unit volume $g(S) \equiv (G(S)-1)/V$,
one finds
for a Gaussian random field \cite{Ha86}:
\be
 g(\nu)= N (1-\nu^{2})\,{\rm e}^{-\nu^{2}/2} .
\ee
Here the amplitude 
\be
\label{Ngauss} 
N=\frac{1}{(2\pi)^2} \left(\frac{\left< k^2\right>}{3}\right)^{3/2} 
\ee
is a constant which depends 
via 
its second moment 
\be
\left< k^2\right>=
\frac{\int{k^2\hat{P}(k)\,{\rm d}^3k}}{\int{\hat{P}(k)\,{\rm d}^3k}} 
\label{k2}
\ee
on the power spectrum $\hat{P}(k)$ 
of the (smoothed) density field.

Hence, the genus curve of a Gaussian random field 
exhibits a universal w-shape. Only
the genus 
amplitude depends on the shape of the power spectrum via equation (\ref{k2}),
but it is independent of the normalization of $\hat{P}(k)$.

For a non-Gaussian density field, different topologies are expected. 
For example, the genus curve can be shifted 
towards a `meatball' topology 
or towards a `Swiss-cheese' topology. In a meatball topology, 
the genus curve peaks at a negative value of $\nu$ because isolated 
structures dominate over an extended range at positive  
densitiies, whereas in a Swiss-cheese topology, the genus curve 
peaks at a positive value of $\nu$, since 
the topology is dominated by empty voids up to a larger density value.

The genus curve defined as a function of $\nu$ has the remarkable 
property that it is invariant during the linear growth of density 
fluctuations in the Universe. The density contrast $\delta \rho /\rho$
grows and contours of $\rho$ change, but contours of $\nu$ are fixed. 
This is a very important property as by measuring the genus curve on
scales where non-linear growth has yet to occur we can recover
information 
about the primordial density field and, eventually, be able to
distinguish 
between Gaussian and non-Gaussian initial conditions.

As an alternative to labeling the isodensity surfaces with the number 
of standard deviations from the mean, it is common to parameterize
them
with the fraction $f_{\rm vol}$ of the survey volume above the given density
threshold or, 
equivalently, by the number $\nu_{\rm f}$ given implicitly by
\be
\label{volfrac}
f_{\rm vol}=\frac{1}{(2\pi)^{1/2}}\int_{\nu_{\rm f}}^{\infty}{\rm e}^{-t^{2}/2}{\rm d}t .
\ee

Note that the two parameterizations coincide for the case of a Gaussian
random field.
The volume fraction parameterization has the disadvantage that it is
less 
sensitive to skewness in the density probability distribution. 
In addition, it does not discriminate between Gaussian distributions
and 
any other distribution which is a one-to-one transformation of the
former, for example, a lognormal distribution \cite{Co91}. 
This is because the contours of constant volume fraction are invariant
under such mappings, as well as the topology of the isodensity
contours themselves.

As Vogeley et~al. \shortcite{Vo94} argue, this invariance property is
indeed desired, if one is 
not interested in the positive
skewness developing during gravitational collapse, 
but rather in the topological properties 
of the initial density field. 
By using the volume fraction to label isodensity contours, the genus
curve becomes independent of 
the one-point probability distribution
function (PDF). In this way the topological analysis is not mixed up
with the one-point PDF, which can be better studied by other means.
Subscribing to this philosophy we will 
hereafter use the volume fraction parameterization.

\subsection{Calculation of the genus curve}

We want to utilize the genus to study the topology of the observed
galaxy distribution, which comes in the form of a point set. 
As a first step we therefore need to adopt a
method to compute suitable surfaces from the point distribution.
Mecke et~al. \shortcite{Me94} 
assigned to each galaxy site a ball of radius $r$ and
examined the genus of the union set of these spheres. We will employ
the more widely used approach 
of constructing smoothed maps
of the PSCz density field (see Section 3)
and considering isodensity contours for the genus statistic.

Our code to calculate the genus curve of a given smoothed density 
field is based on the algorithm first proposed by 
Gott et~al.\ \shortcite{Go86} and on the 
{\small CONTOUR} code by Weinberg \shortcite{We88}. 
We have written a new implementation in C, 
able to compute high resolution genus curves very quickly. 
Here a major increase in speed was achieved by sorting the discrete density
field first in order to be able to 
find the threshold values $\rho_{\rm t}$ for the
required values of $f_{\rm vol}$ at no computational expense.

Given a particular surface of constant density, we approximate it by 
a network of polygonal faces. This procedure does not change the
global topology of the surface as long as the grid size is much
smaller 
than the smoothing length. 
    
When such a polygon network is used to approximate a compact surface
of 
genus $G$, it can be shown \cite{Go86} that
the genus is given by
\be
\sum_{i}D_{i}=4\pi\,(1-G) ,
\ee
where $D_{i}=360^{\circ}-\sum_{i}V_{i}$ is the angle deficit at 
each vertex and $V_{i}$ are the angles around the vertex. For example,
in a cube, 
which approximates a compact surface of genus 0, there are  
three squares around each vertex and, thus, 
$D_{i}=360^{\circ} - 3\times 90^{\circ}=90^{\circ}$ 
at each of the eight vertices, giving
$\sum_{i}D_{i}=720^{\circ}=4\pi$, 
as expected. The reason for this is that the curvature is 
compressed 
into $\delta$-functions at the vertices. 
Parallel transport arguments show that the integral of the 
$\delta$-functions of Gaussian curvature over the infinitesimal 
area of a vertex is just equal to the angle deficit at that vertex.

In practice we 
divide space into cubic lattices for sampling either 
galaxy or simulation data; we then 
calculate the genus of a given isodensity surface by 
approximating it by a network of square faces, 
and by adding up the angle 
deficits of all vertices.

\subsection{Genus related statistics}

In order to measure departures of the observed 
genus curve from the random-phase shape we use 
the genus meta-statistics introduced by Vogeley
et~al.\ \shortcite{Vo94}. They consist of an appropriately defined
amplitude, width, and shift of the genus curve.
Additionally we consider the {\it amplitude drop} of the genus curve
compared to the random phase expectation.

\subsubsection{Amplitude}

We measure the amplitude of the genus curve as the amplitude $N$ of
the best fit random phase genus curve by minimizing $\chi^2$ 
in the range $-1< \nu < 1$. 
If the underlying density field is sufficiently close to Gaussian the
amplitude $N$ provides direct information on the shape of the power
spectrum.
It should be noted, however,
that 
the amplitude 
is systematically biased high by shot noise due to finite 
sampling of the density field, as will be discussed below.

\subsubsection{Width}

The second of the meta-statistics is the width of the genus curve, 
which refers to the 
range of $\nu$ over which the genus is positive, i.e.
\be 
W_{\nu}=\nu_{+}-\nu_{-},
\ee
where $\nu_{+}$ and $\nu_{-}$ are the first zero 
crossings of the genus curve, right and left of the origin. 
The width $W_{\nu}$ measures 
whether the examined
density field is more or less sponge-like than a random-phase
field, which has $W_{\nu}=2$.
In order to improve the reliability of the determination of  $\nu_{+}$
and $\nu_{-}$
we first reduce the noise in the genus curve by boxcar
smoothing it with a filter of total width $\Delta\nu=0.2$. This
smoothing is only applied for this statistic.

\subsubsection{Shift}

In order to distinguish between a Swiss-cheese and a 
meatball topology, we use a measure for the shift of 
the peak of the genus curve.
This shift may be quantified via
\be
\Delta\nu=\frac{\int{\nu \,g(\nu)_{\rm obs}{\rm d}\nu}}
{\int{g(\nu)_{\rm fit}{\rm d}\nu}} ,
\ee
where $g(\nu)_{\rm obs}$ is the observed genus curve and $g(\nu)_{\rm fit}$ is
the corresponding 
best-fit random-phase curve \cite{Pa92a}. We evaluate $\Delta\nu$ in the range  $-1<\nu<1$. 
A value of $\Delta\nu>0$ indicates a bubble-shift, whereas $\Delta\nu<0$ means a
meatball-shift.

\subsubsection{Amplitude drop\label{drop}}

With the above meta-statistics,
any positive detection
of a deviation of the genus curve 
from the random phase shape 
signals the presence of non-Gaussian
features in the density field. However, the converse is not
necessarily true. 
For example, we find for the N-body models analysed below that the
genus curve retains its random phase shape almost perfectly at all
smoothing scales, even when
the smoothed density field exhibits already strong non-Gaussian
features that develop in the non-linear growth of structure. However,
the amplitude of these genus curves is reduced compared
to the expectation based on the power spectrum alone. 

This amplitude drop can be
taken as evidence for phase correlations that 
develop in the 
weakly non-linear regime. 
In order to quantify this effect, we 
define the 
amplitude drop 
\be
R=\frac{N}{N_{\rm rp}},
\ee
where $N_{\rm rp}$ is the
amplitude 
of the random-phase density field with the same power
spectrum as the examined density field. 

There are different
possible ways to measure $N_{\rm rp}$. 
For a fully sampled N-body
simulation 
with periodic boundary conditions, perhaps the simplest way is to
Fourier 
transform the original density field, randomize the phases in 
Fourier space subject to the condition ${\delta_{k}}^*=\delta_{-k}$, 
therefore {\it Gaussianizing} the field, and then transform back to 
real space and measure the genus again. 
This procedure will be applied to 
the simulations, that we study in Section 5. 

For the case of PSCz, there are 
non-trivial phase correlations between the mask 
and the density field due to the odd shaped 
survey volume, which spoil this procedure.
Instead of constructing a representation of the corresponding random phase
density field, the estimate of $N_{\rm rp}$ needs to be based on a
direct measurement of the second moment of the power spectrum.

For this purpose, Vogeley et~al.\ \shortcite{Vo94}
estimated the power spectrum directly, but this is
also complicated by
the presence of the mask, and in addition it requires a subtle integration of
the measured spectrum to derive $N_{\rm rp}$.

As an alternative we propose to measure the variance
$\sigma_{\rm t}^2(\lambda)$ of the underlying density field as a function of 
smoothing length. This allows an estimate of $\left< k^2\right>$ via
the logarithmic slope of $\sigma_{\rm t}^2(\lambda)$:
\be
\left<k^2\right>\lambda^{2}=-\frac{d\log\sigma_{\rm t}^2}{d\log\lambda}.
\label{eqslope}
\ee
From this the amplitude 
$N_{\rm rp}$
can be obtained using equation (\ref{Ngauss}). Our estimator for the
variance $\sigma_{\rm t}^2(\lambda)$ is discussed in Appendix \ref{appvar}. 
In order to reduce the cosmic variance in the estimate of
$\left<k^2\right>\lambda^{2}$
we actually fit the measured variances with 
an analytic form based on a generic CDM power spectrum, and compute
the derivative from this fit.

\subsection{Error estimates}

\subsubsection{Uncertainty due to finite sampling}

We assume that the observed galaxy distribution represents a 
Poisson realization of the underlying density field. Since we 
have
only one sampling realization of the PSCz density field at our
disposal we employ the bootstrap technique to derive estimates of the
uncertainties due to sampling noise in the PSCz genus results.

For this purpose we generate an ensemble of 
bootstrap resamplings of the PSCz galaxy
catalogue by randomly redistributing the galaxies onto the sites
provided by the original catalogue. Note that certain positions will be
left empty while others will be occupied by several galaxies.
An estimate for the 
statistical uncertainty of a measured quantity due to finite sampling 
may then be obtained as the variance among the measurements for the bootstrap
ensemble.

In the context of genus statistics this method has been proven to be 
a reliable and robust -- if somewhat conservative -- estimator of
statistical 
errors. 
Moore et~al.\ \shortcite{Mo92} compared bootstrap errors for the 
genus statistic with the statistical errors obtained for different 
resamplings of full N-body simulations and found that the bootstrap 
errors are 10 per cent higher than the `true' statistical uncertainty.

Of course, in addition to the sampling noise the genus curves are
also uncertain due to cosmic variance resulting from the finite survey
volume accessible to PSCz. 
We believe that PSCz is the best sample examined so far in this
respect and that due to the large number of resolved structure
elements the cosmic variance is smaller than in previous studies.
However, perhaps the only reliable 
way to quantify its influence is to examine PSCz-like mock
surveys extracted from large-volume N-body simulations. 
This technique allows a careful statistical study 
of systematic
effects and will be applied in a forthcoming paper by
Springel et~al.\ \shortcite{Sp97b}.

\subsubsection{Shot noise bias}

In addition to the variance introduced by finite sampling there is also 
a systematic effect on the genus curve due to shot 
noise. 

In Appendix \ref{appshot} we calculate an estimate of this bias. 
To the extent 
that the shot noise does not spoil the Gaussian nature of the
underlying field, the genus amplitude may be estimated by
\be
\label{Nundershot}
N=\frac{1}{(2\pi)^2}\left[\frac{\left< k^2\right> }{3}+\frac{\sigma^2_{\rm shot}}{\sigma^2_{\rm t}+\sigma^2_{\rm shot}}\left(\frac{1}{\lambda^2}-\frac{\left< k^2\right> }{3}\right)\right]^{3/2}.
\ee
Here, $\sigma^2_{\rm shot}$ is the variance introduced 
by shot noise on a given scale, and 
$\sigma^2_{\rm t}$ is the variance in the field free from shot noise 
on that same scale.

When the power spectrum follows
a power law with slope $n$
on the relevant scales, 
the amplitude becomes 
\be
N=\frac{1}{(2\pi)^2\lambda^3}\left[\frac{n+3}{3}+\frac{\sigma^2_{\rm shot}}{\sigma^2_{\rm t}+\sigma^2_{\rm shot}}\left( \frac{-n}{3}\right) \right]^{3/2}.
\ee

We tested this relation with Monte-Carlo experiments 
and found it to 
be fulfilled very well.
Since in a realistic application we typically have $-3<n<0$,
shot-noise tends to increase the measured genus amplitude.

In order to quantify the contribution of shot noise to the 
genus amplitude we introduce the ratio
\be
A_{\rm shot}=\frac{N_{\rm obs}}{N_{\rm t}} ,
\ee
where $N_{\rm obs}$ is the observed genus amplitude under the 
influence of noise and  $N_{\rm t}$ is the amplitude of the field 
free from shot noise. 

Since the variance of the density field due to shot noise 
$\sigma_{\rm shot}$ varies with distance, we compute an
estimate of $A_{\rm shot}$ by 
averaging 
relation (\ref{Nundershot}) over the survey volume.

The variances $\sigma^2_{\rm t}$ and $\sigma^2_{\rm shot}(r)$ 
are estimated as outlined in Appendix \ref{appvar}. We also need 
the 
logarithmic slope $\left<k^2\right>\lambda^{2}$ 
defined in equation (\ref{eqslope}).
In practice, we want to obtain an estimate for $A_{\rm shot}$ that is
not strongly affected by fluctuations due to cosmic variance between
the different adopted survey volumes. We therefore base the 
estimate of $A_{\rm shot}$ 
on values for $\sigma^2_{\rm t}$ derived from an analytic
fit to the measured variances, and on slopes
$\left<k^2\right>\lambda^{2}$ obtained for this fit.

With these quantities computed, 
the integrations required to estimate $A_{\rm shot}$ can be done. 
Typically, we find $A_{\rm shot}\sim 1.1 - 1.2$ for PSCz, 
so that genus amplitudes 
are increased by up to 20 per cent. It is therefore crucial to correct 
amplitudes for shot noise if one is to draw quantitative conclusions 
from the measured genus amplitude.
For this reason, we
calculate a corrected amplitude $N^*=N/A_{\rm shot}$ in 
order to estimate the amplitude drop $R$, which is particularly
sensitive to the value of the amplitude.

\section{Construction of Density Maps}

        \label{denmaps}
	
\subsection{The PSCz redshift survey data}

\begin{figure}
\bc
\resizebox{8cm}{!}{\includegraphics{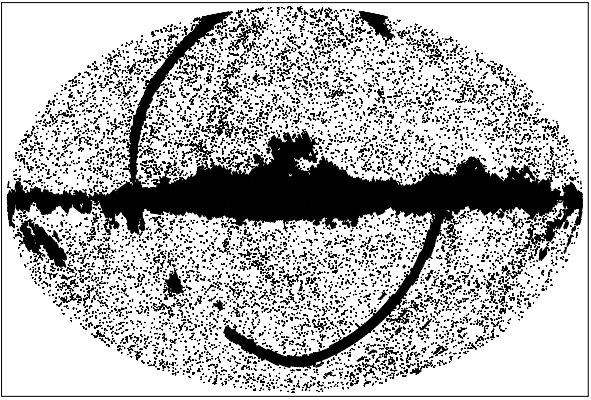}}
\caption
{Sky coverage of the PSCz survey in an Aitoff projection. The dots
represent the galaxies in the survey, the shaded regions are
unobserved and comprise the angular mask. The galactic center lies in
the center of the plot.
\label{figMask}}
\ec
\end{figure}

The sky coverage of the PSCz redshift survey is 84.1 per cent, excluding only 
the zone of avoidance, here defined by an 
infrared background exceeding 25$\,$MJy$\,$sr$^{-1}$ at 100$\,\mu$m, and 
a few unobserved or contaminated patches at higher latitude.  
The excluded regions are coded in an angular mask as shown in Figure 
\ref{figMask}.

We convert the observed redshifts to the Local Group frame and use them
directly to infer comoving distances without further corrections for peculiar
velocities.
As has been demonstrated with N-body
experiments \cite{Pr97} and analytical calculations \cite{Ma96b},
redshift space distortions hardly affect genus
statistics. Therefore this should be a good approximation for the genus
analysis. However, a few blue-shifted galaxies had to be discarded, resulting in an effective sample of 14505 galaxies.
We will assume an Einstein-de-Sitter model for the
background cosmology throughout. The results should not be sensitive to
this choice because in a cosmological context the PSCz density field
maps only the local Universe.

\subsection{Smoothing procedure}

Assuming a universal luminosity function, an unbiased estimate of 
the galaxy density field $\rho(r)$ can
be obtained by weighting the discrete point distribution $m(\vec{r})$ 
of the
observed galaxies with the inverse of the selection function $S(r)$:
\be
\rho(\vec{r}) \propto \frac{m(\vec{r})}{S(r)}.
\label{EQ0}
\ee
Here the selection function $S(z)=\left<m(\vec{r})\right>$ is defined
as the mean expected comoving number density of sources at redshift
$z$ corresponding to the comoving position $\vec{r}$.
 We employ the fitting form
\be
S(z)=\frac{\psi}{z^{\alpha}\left(
1+\left(\frac{z}{z^{\star}}\right)^{\gamma}\right)^{\beta/\gamma}}\; ,
\label{EQ10}
\ee
and determine its parameters (see Table \ref{tab1})
with the methods outlined in
Springel \& White \shortcite{Sp97a}. 
Note that the selection function
includes a correction for the strong evolution seen in \iras galaxies.

In order to obtain an estimate of the density field smoothed on some scale
$\lambda$ we convolve $\rho(\vec{r})$ with a Gaussian filter of the form 
\be
W(\vec{x})=\frac{1}{\pi^{3/2}\lambda^{3}}\,\exp\left({-\frac{\vec{x}^{2}}{\lambda^{2}}}\right).
\label{EQ1}
\ee
Note that here we use Gott et~al.'s \shortcite{Go89} 
definition of a Gaussian filter rather 
than the conventional form.

However, due to the lack of galaxies in the regions 
of the angular mask, the density would be systematically 
underestimated at locations close to unobserved patches 
if the smoothing were just done by a straightforward
use of the kernel of equation (\ref{EQ1}).
In order to avoid this problem we employ the ratio method proposed by Melott \&
Dominik \shortcite{Me93}, who 
have shown in a systematic study that a smoothing according to 
\be
\hat{\rho}(\vec{r})=\frac{\int W(\vec{r}-\vec{r}') \rho(\vec{r}')
\,\dd\vec{r}'}
{\int W(\vec{r}-\vec{r}'') M(\vec{r}'')
\,\dd\vec{r}''},
\label{EQ2}
\ee
leads to the smallest loss or distortion of topological information
compared to a number of alternative schemes that treat the
mask differently. Here $M(\vec{r})$ is a mask field 
defined to be equal to 0 for
$\vec{r}$ lying behind the angular mask and to be 1 otherwise. 
For this choice the 
denominator of equation (\ref{EQ2}) essentially renormalizes the
smoothing kernel to the survey volume visible from the reference point
$\vec{r}$.

In the actual 
computation of the genus curve we only 
use the volume with $M(\vec{r})=1$ which is not hidden by the
mask. Additionally we restrict the genus computation to a sphere
carved out of the smoothed density field. Note that there is no boundary
smoothing effect due to the outer surface of this sphere since we also
include the sources outside this final region in the smoothing
process.

The method we apply here has the advantage that it does not
require some form of {\em filling} 
of the unobserved regions. As Melott \& Dominik \shortcite{Me93}
have shown, simple forms of such fillings that come to mind,
like a constant density padding or randomly placed points of the mean
background density, lead to larger biases in the genus curve than the
ratio method.

We compute the convolutions that appear in the numerator and denominator of
(\ref{EQ2}) using a Fast Fourier Transform (FFT) on a
 $128^3$ mesh. We choose a grid size of $b=\lambda/6$, which
ensures that the genus is free from finite mesh size
effects \cite{Ha86}. 
The final depth $R_{\rm max}$ of the 
density field that we use for the topological analysis
is always small enough to avoid wrap around effects due to the periodic
FFT smoothing.

\subsection{Depth of maps}

Because the galaxy density of a flux limited sample declines quickly
with distance, the uncertainty of the smoothed density estimate grows
rapidly with redshift. It is desirable, of course, to use a
survey volume that is as large as possible 
in order to beat down statistical noise and cosmic variance.
However, with respect to the genus statistics the sampling must be at
least dense enough to make
discreteness effects in the genus curve negligible. 

According to a useful rule of thumb \cite{We87} 
discreteness effects are small  
if 
\be
\lambda\ge d=S^{\,-\frac{1}{3}} \, ,
\label{C15}
\ee
where $d$ is the mean interparticle separation and $S$ is the selection function. 
Adopting this criterion we choose a
maximal radius $R_{\rm max}$ 
given by 
$\lambda= S(R_{\rm max})^{-\frac{1}{3}}$
and use it to delimit 
the usable survey volume $V_{\rm s}$. This choice ensures that at the far
edge of the survey volume the sampling condition is just met, and in the
remainder of the volume the sampling is denser.

However, if $\lambda$ is of the order of $d$ one risks artificially
introducing a meatball bias by identifying lonely
tracers as isolated clusters. Therefore, 
any observed bias of this kind should be treated with caution, since it
might well be a discreteness effect.

Additionally, the genus amplitude is biased due to  
shot-noise present in the reconstructed density field, as we have
discussed in Section 2.
This noise contributes to second and higher moments of the density field.
Unfortunately, the genus depends in a non-linear way on the various
moments of the density field, making it extremely difficult to 
derive an unbiased genus estimator which is not affected by it.
Hence, we estimate the contribution
of shot noise to the genus signal {\em a posteriori}.

\begin{figure}
\bc
\resizebox{8cm}{!}{\includegraphics{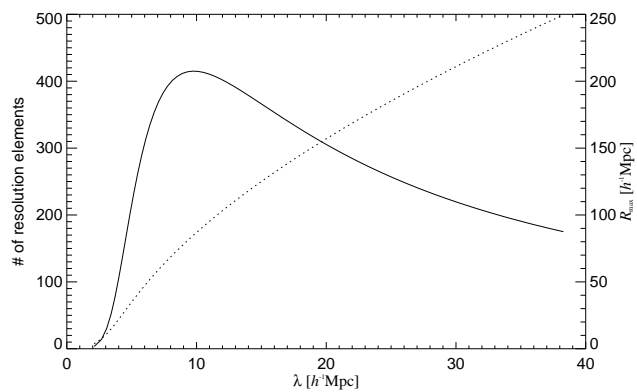}}
\caption
{The number of resolution elements (solid, left hand scale) for the PSCz survey when the maximal
survey volume is used. Also shown is the radius (dashed, right hand scale) of the 
usable survey volume.\label{fig1}}
\ec
\end{figure}

\subsection{Resolution elements}

The notion of {\it number of resolution elements} provides 
a useful way to roughly compare the statistical power of genus
measurements.
Because the 
smoothing extends over an effective volume
$V_{\rm sm}=\pi^{3/2}\lambda^3$
the number of independent structures that can be present in a finite
survey volume is limited. This number is of order
\be
N_{\rm res}=\frac{V_{\rm s}}{V_{\rm sm}}=\frac{ \omega
R_{\rm max}^3}{3\pi^{3/2}\lambda^3},
\ee
where $\omega$ is the solid angle
covered by the survey.

The number $N_{\rm res}$ indicates the power of a data set used for
topological analysis.
With the QDOT survey Moore et~al.\ \shortcite{Mo92} reached a maximum of about
$N_{\rm res} =80$ whereas the most powerful genus results so far
came from Vogeley et~al.'s \shortcite{Vo94} treatment of the CfA
survey, where they
reached $N_{\rm res}=260$ for their best subsample. In a recent study
of the 1.2-Jy redshift survey Protogeros \& Weinberg \shortcite{Pr97}
achieved $N_{\rm res} = 170$.

The PSCz redshift survey can provide still more resolution elements as
is evidenced in Figure \ref{fig1} and Table \ref{tab2}. 
It also provides a high number of
resolution elements over an unprecedented wide range of smoothing
scales. In particular, there are more than $300$ resolution elements
in the range $6\lu\le \lambda \le 20\lu$. This wide dynamic range
together with the large volume covered make it more powerful than all
previously examined samples.

For the genus analysis we have examined smoothing lengths
between $5\lu$ and $56\lu$ with approximately logarithmic spacing. In
particular, all the smoothing lengths considered in the studies by
Moore et~al.\ \shortcite{Mo92}, Vogeley et~al.\ \shortcite{Vo94} 
and Rhoads et~al.\ \shortcite{Rh94} 
are contained in this set.
Table \ref{tab2} lists some relevant parameters for the different
cases.

\begin{table}
\bc
\caption{Parameters of the selection function of PSCz.\label{tab1}}
\begin{tabular}{ccc}
$\alpha$  & $\beta$ & $\gamma$ \\
$0.991^{+0.068}_{-0.073}$  
&  $3.445^{+0.173}_{-0.158}$ 
& $1.925^{+0.162}_{-0.153}$ \\
\\
$z^{\star}$ & $\psi \;\;[h^3{\rm Mpc}^{-3}]$ & \\
$0.02534^{+0.00130}_{-0.00116}$ 
&$(141.3\pm 2.4 ) \times 10^{-6}$ & \\
\end{tabular}
\ec
\end{table}

\begin{table}
\bc
\caption{The smoothing lengths adopted for the topological analysis of the PSCz
survey. Listed are the adopted survey depth $R_{\rm max}$, the
resulting number
$N_{\rm res}$ of resolution elements and the number $N_{\rm gal}$ 
of galaxies inside the survey volume.
\label{tab2}
}
\begin{tabular}{c|c|r|r|r}
$\lambda\;\;[\lu]$  & $R_{\rm max}\;\;[\lu]$  &
$N_{\rm res}$ & $N_{\rm gal}$  \\
 5&  34.92&  215.5&    2295\\
 6&  47.16&  307.3&    3550\\
 7&  58.35&  366.3&    4928\\
 8&  68.58&  398.5&    5909\\
10&  86.91&  415.3&    7510\\
12& 103.19&  402.3&    8775\\
14& 118.04&  379.2&    9681\\
16& 131.81&  353.7&   10356\\
20& 156.96&  305.8&   11309\\
24& 179.79&  266.0&   11968\\
28& 200.97&  233.9&   12496\\
32& 221.79&  210.6&   12833\\
40& 263.44&  180.7&   13339\\
48& 305.09&  162.4&   13684\\
50& 315.50&  158.9&   13748\\
56& 346.73&  150.2&   13924\\
\end{tabular}
\ec
\end{table}

\section{PSCz results}

        \label{results}		
        
\subsection{PSCz genus curves}

\begin{table*}
\bc
\caption{Summary of numerical results obtained for the genus
meta-statistics of PSCz.
Listed for each smoothing length are the genus amplitude in
the dimensionless form $(2\pi)^2\lambda^3 N$, the absolute
genus amplitude $G=\frac{4\pi}{3} N R_{\rm max}^3 $ 
in the survey volume, the shift $\Delta\nu$,
and the width $W_{\nu}$. We also give the measured variance $\sigma^2$
of the smoothed density fields, and our estimated shot-noise
correction factors $A_{\rm shot}$ and the amplitude drop $R$.
}
\begin{tabular}{cccrcccc}
$\lambda\,\,[h^{-1}{\rm Mpc}]$ &  $(2\pi)^2\lambda^3 N$ & $G$ & 
$\Delta\nu\:\:\:$ & $W_\nu$ & $\sigma^2$ & $A_{\rm shot}$ & $R$ \vspace{0.2cm}\\
           5 & $ 0.20\pm  0.03$ & $  7.40\pm  1.12$ & $-0.24\pm  0.12$ & $ 2.24\pm  0.20$ & $ 0.6691\pm  0.1312$ & $ 1.09$ & $ 0.58\pm  0.09$ \\ 
           6 & $ 0.31\pm  0.03$ & $ 16.05\pm  1.44$ & $-0.14\pm  0.07$ & $ 2.01\pm  0.10$ & $ 0.6044\pm  0.0772$ & $ 1.10$ & $ 0.81\pm  0.07$ \\ 
           7 & $ 0.36\pm  0.04$ & $ 22.05\pm  2.38$ & $-0.04\pm  0.07$ & $ 2.13\pm  0.12$ & $ 0.5163\pm  0.0485$ & $ 1.10$ & $ 0.87\pm  0.09$ \\ 
           8 & $ 0.41\pm  0.02$ & $ 27.11\pm  1.23$ & $ 0.05\pm  0.05$ & $ 2.19\pm  0.17$ & $ 0.3937\pm  0.0315$ & $ 1.11$ & $ 0.92\pm  0.04$ \\ 
          10 & $ 0.45\pm  0.03$ & $ 31.20\pm  1.99$ & $ 0.05\pm  0.03$ & $ 2.16\pm  0.08$ & $ 0.2463\pm  0.0170$ & $ 1.13$ & $ 0.92\pm  0.06$ \\ 
          12 & $ 0.47\pm  0.04$ & $ 31.58\pm  2.52$ & $ 0.02\pm  0.05$ & $ 2.14\pm  0.09$ & $ 0.1873\pm  0.0201$ & $ 1.14$ & $ 0.88\pm  0.07$ \\ 
          14 & $ 0.48\pm  0.03$ & $ 30.43\pm  2.18$ & $-0.05\pm  0.05$ & $ 2.11\pm  0.10$ & $ 0.1474\pm  0.0156$ & $ 1.16$ & $ 0.84\pm  0.06$ \\ 
          16 & $ 0.51\pm  0.05$ & $ 30.44\pm  3.19$ & $-0.03\pm  0.04$ & $ 2.28\pm  0.10$ & $ 0.1162\pm  0.0122$ & $ 1.17$ & $ 0.84\pm  0.09$ \\ 
          20 & $ 0.64\pm  0.05$ & $ 32.75\pm  2.73$ & $ 0.05\pm  0.06$ & $ 2.08\pm  0.14$ & $ 0.0751\pm  0.0079$ & $ 1.19$ & $ 0.94\pm  0.08$ \\ 
          24 & $ 0.83\pm  0.06$ & $ 36.89\pm  2.59$ & $ 0.04\pm  0.05$ & $ 1.86\pm  0.12$ & $ 0.0482\pm  0.0055$ & $ 1.20$ & $ 1.12\pm  0.08$ \\ 
          28 & $ 0.83\pm  0.07$ & $ 32.73\pm  2.59$ & $ 0.07\pm  0.07$ & $ 1.94\pm  0.18$ & $ 0.0336\pm  0.0042$ & $ 1.20$ & $ 1.06\pm  0.08$ \\ 
          32 & $ 0.79\pm  0.09$ & $ 27.97\pm  3.19$ & $ 0.03\pm  0.07$ & $ 1.81\pm  0.14$ & $ 0.0239\pm  0.0033$ & $ 1.20$ & $ 0.95\pm  0.11$ \\ 
          40 & $ 0.71\pm  0.09$ & $ 21.53\pm  2.68$ & $-0.06\pm  0.06$ & $ 1.95\pm  0.22$ & $ 0.0145\pm  0.0023$ & $ 1.18$ & $ 0.79\pm  0.10$ \\ 
          48 & $ 0.61\pm  0.08$ & $ 16.54\pm  2.16$ & $-0.01\pm  0.08$ & $ 2.11\pm  0.20$ & $ 0.0103\pm  0.0019$ & $ 1.16$ & $ 0.64\pm  0.08$ \\ 
          50 & $ 0.68\pm  0.11$ & $ 18.22\pm  2.93$ & $ 0.03\pm  0.09$ & $ 2.06\pm  0.20$ & $ 0.0100\pm  0.0019$ & $ 1.15$ & $ 0.72\pm  0.12$ \\ 
          56 & $ 0.66\pm  0.11$ & $ 16.65\pm  2.83$ & $ 0.06\pm  0.09$ & $ 2.09\pm  0.18$ & $ 0.0089\pm  0.0018$ & $ 1.12$ & $ 0.68\pm  0.12$ \\ 

\end{tabular}
\ec
\end{table*}

\begin{figure*}
\bc
\resizebox{8cm}{!}{\includegraphics{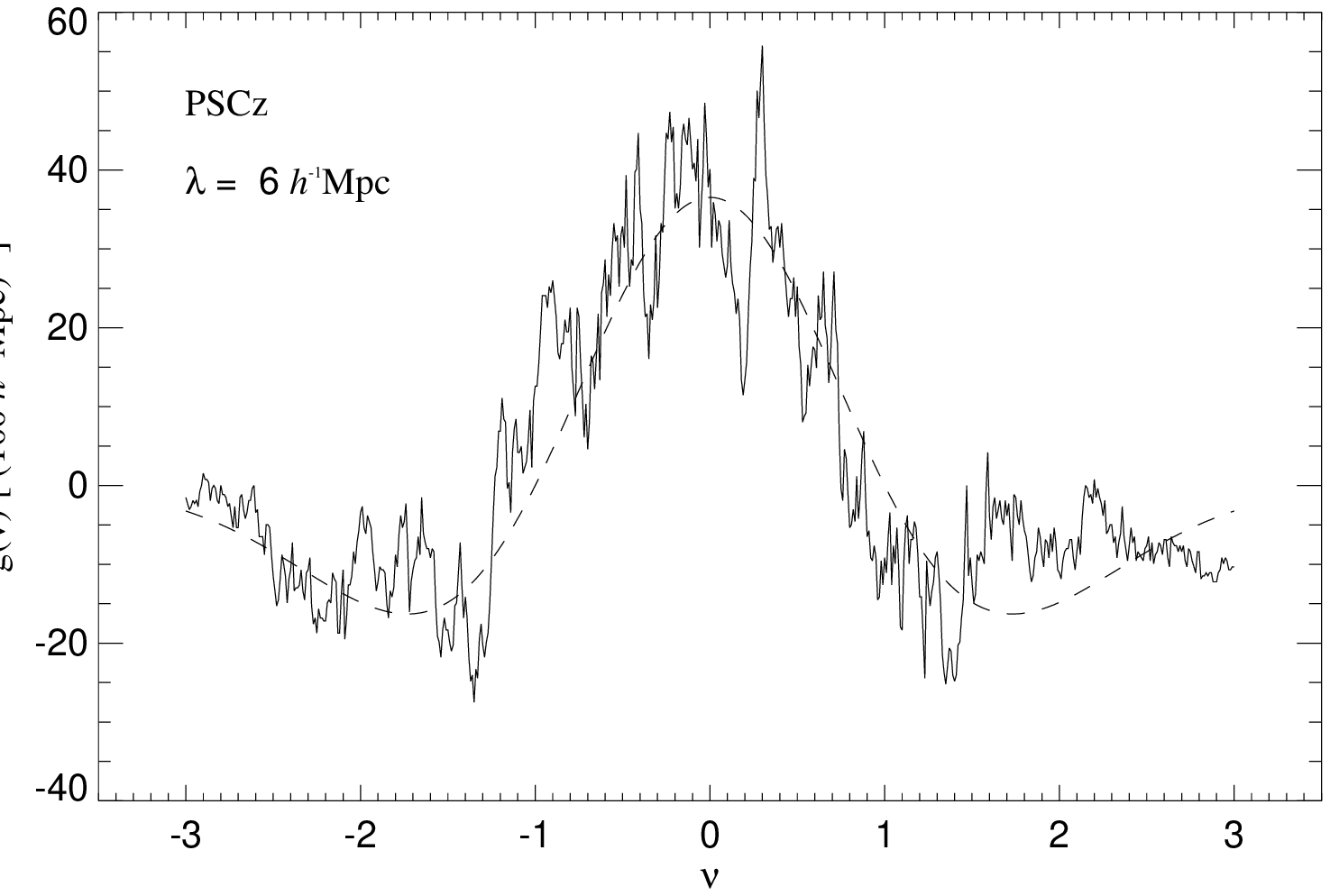}}
\resizebox{8cm}{!}{\includegraphics{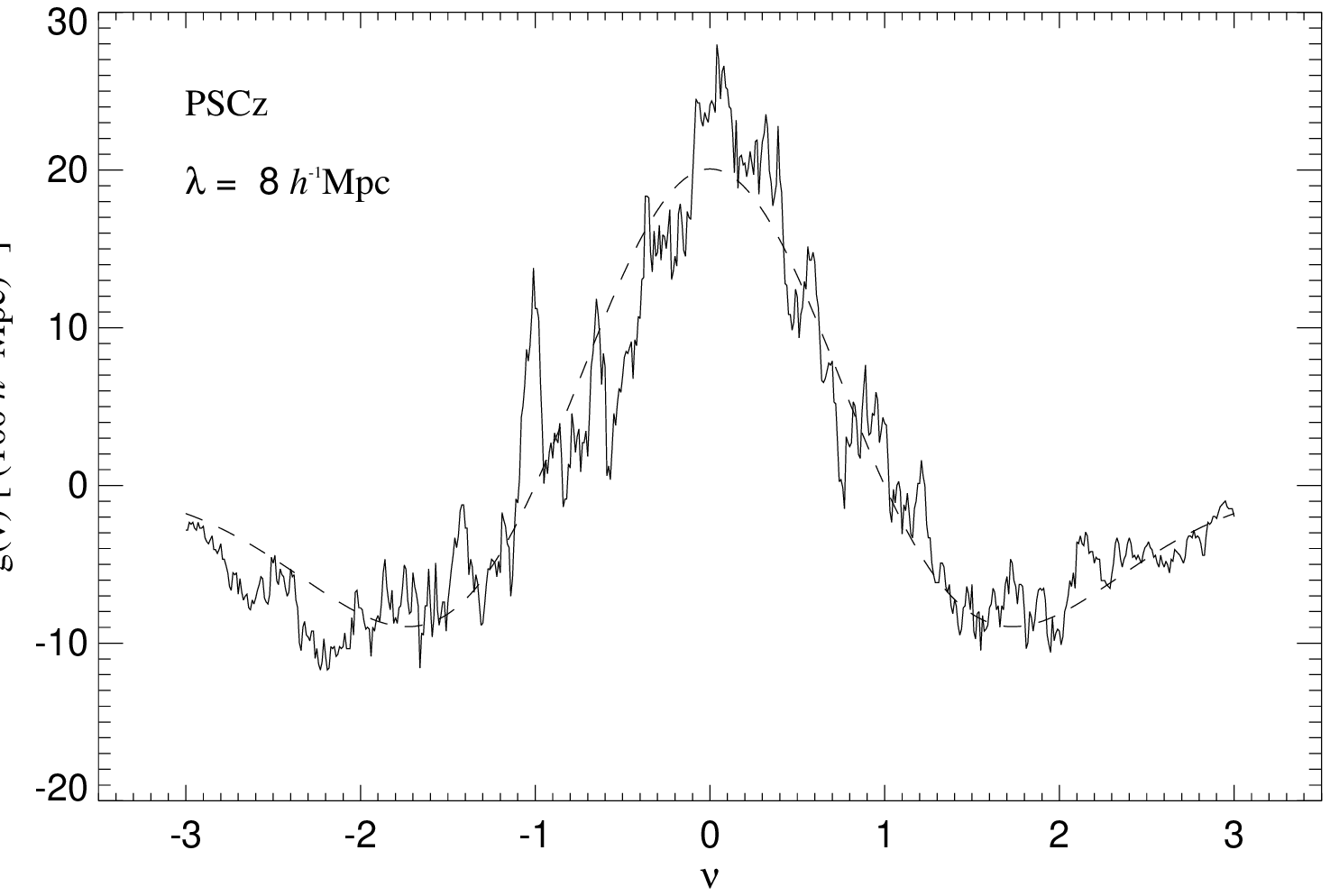}}
\resizebox{8cm}{!}{\includegraphics{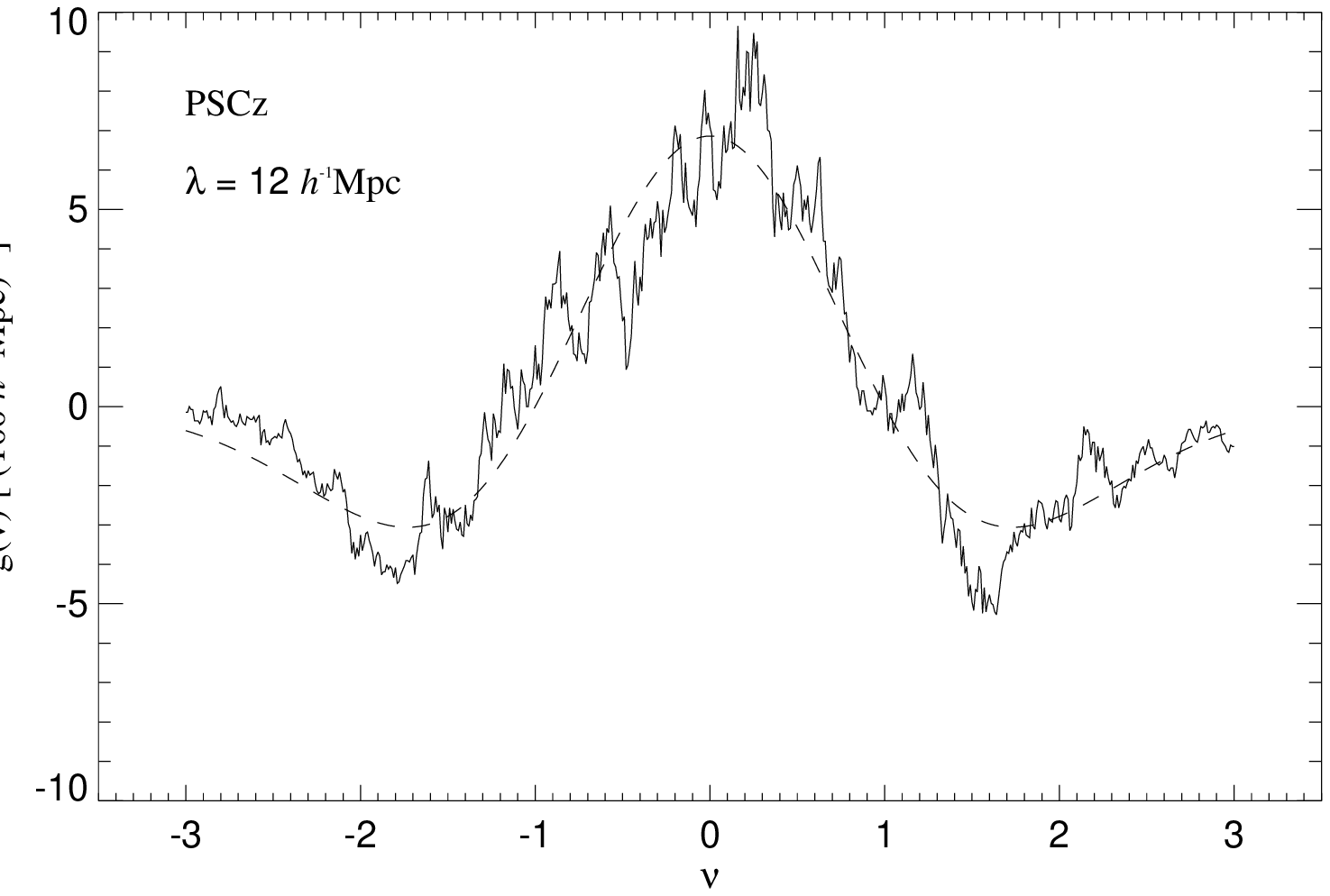}}
\resizebox{8cm}{!}{\includegraphics{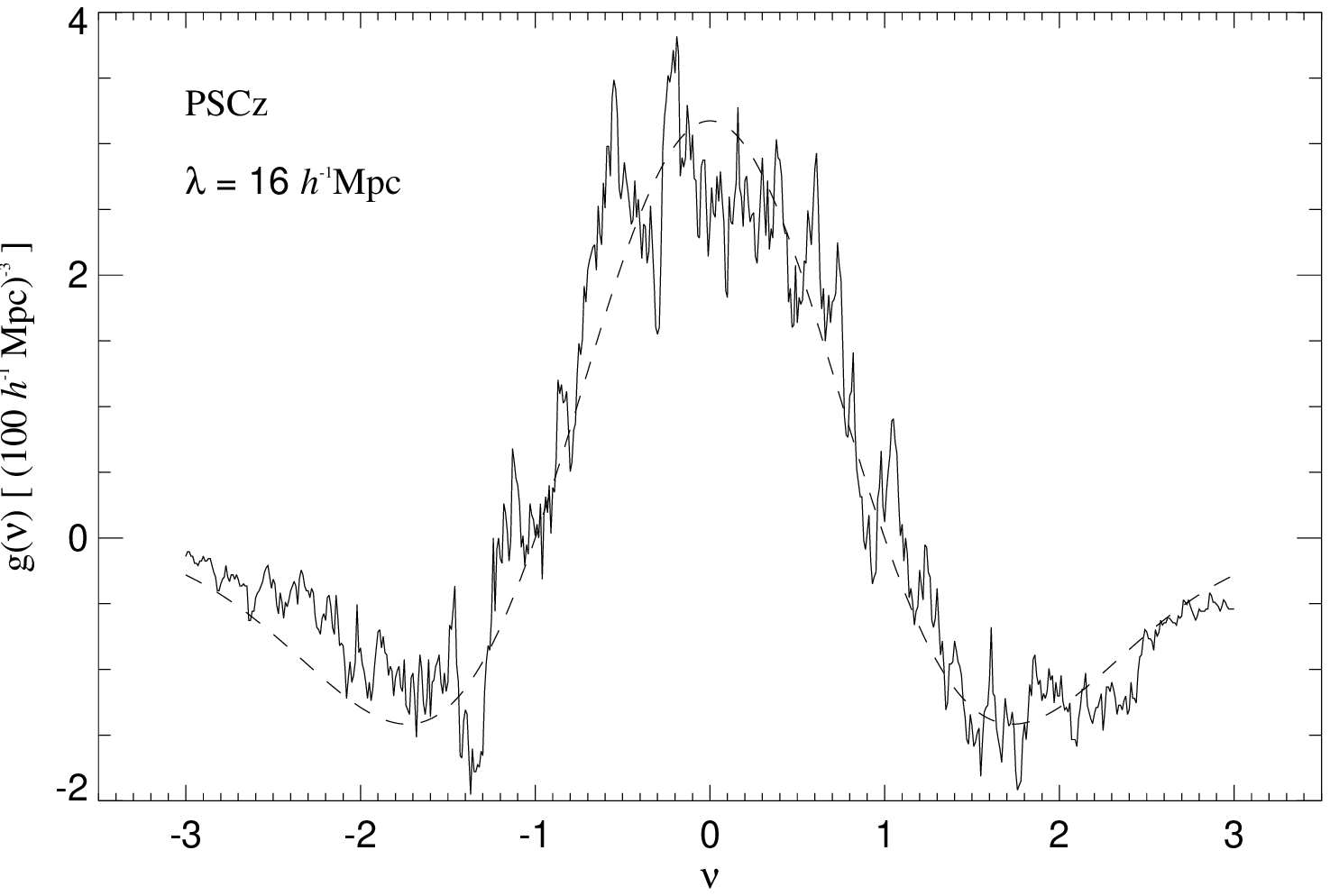}}
\resizebox{8cm}{!}{\includegraphics{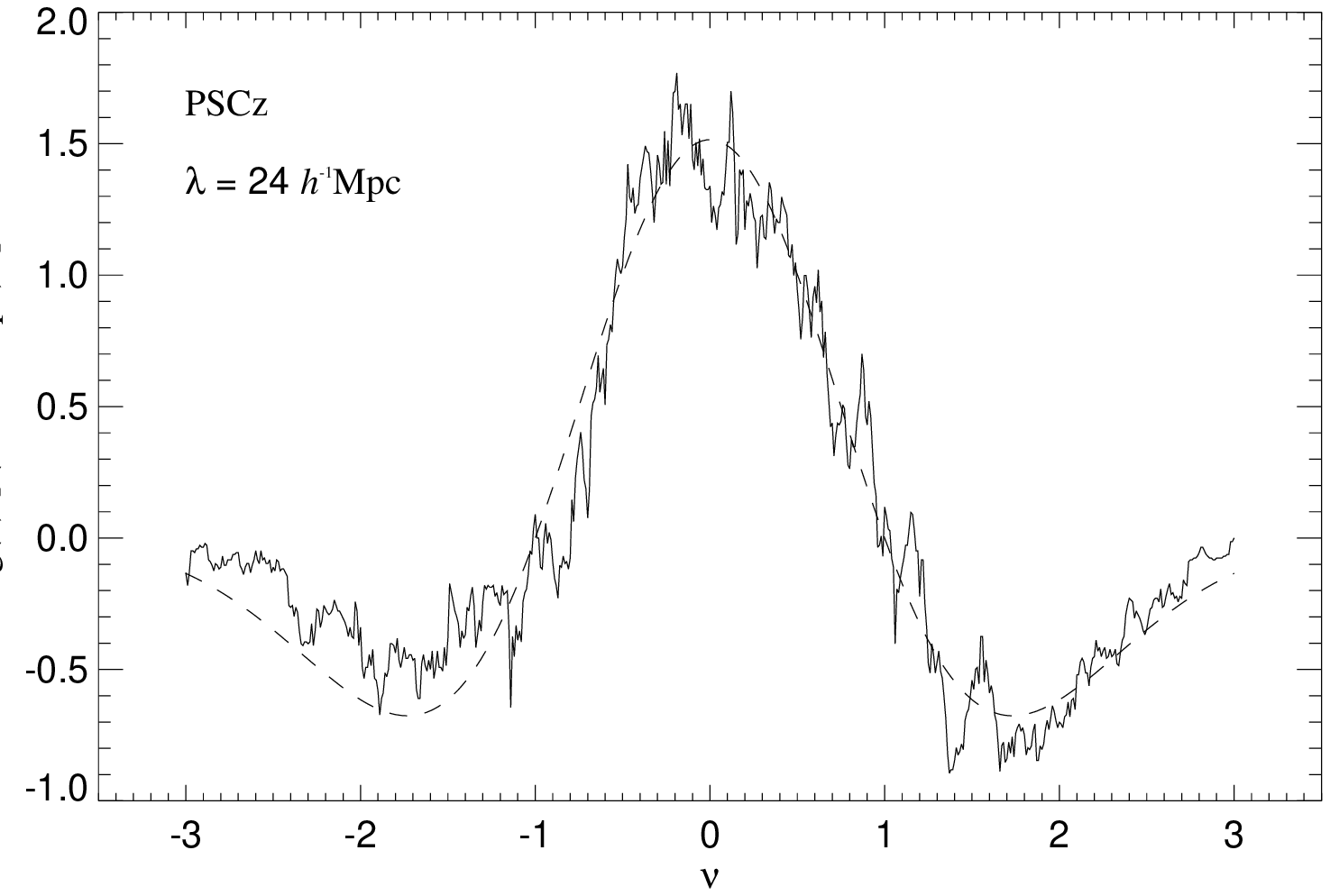}}
\resizebox{8cm}{!}{\includegraphics{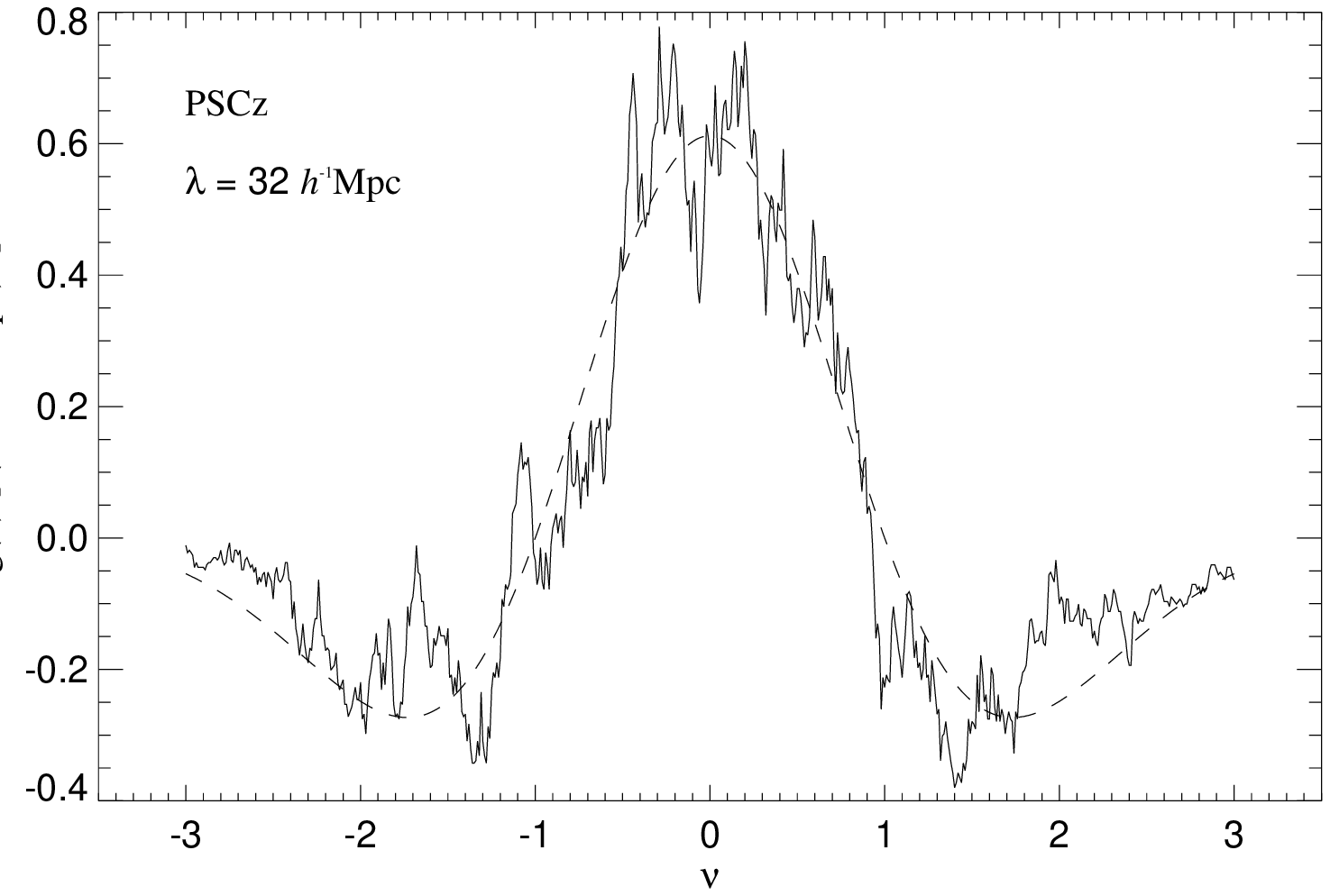}}
\resizebox{8cm}{!}{\includegraphics{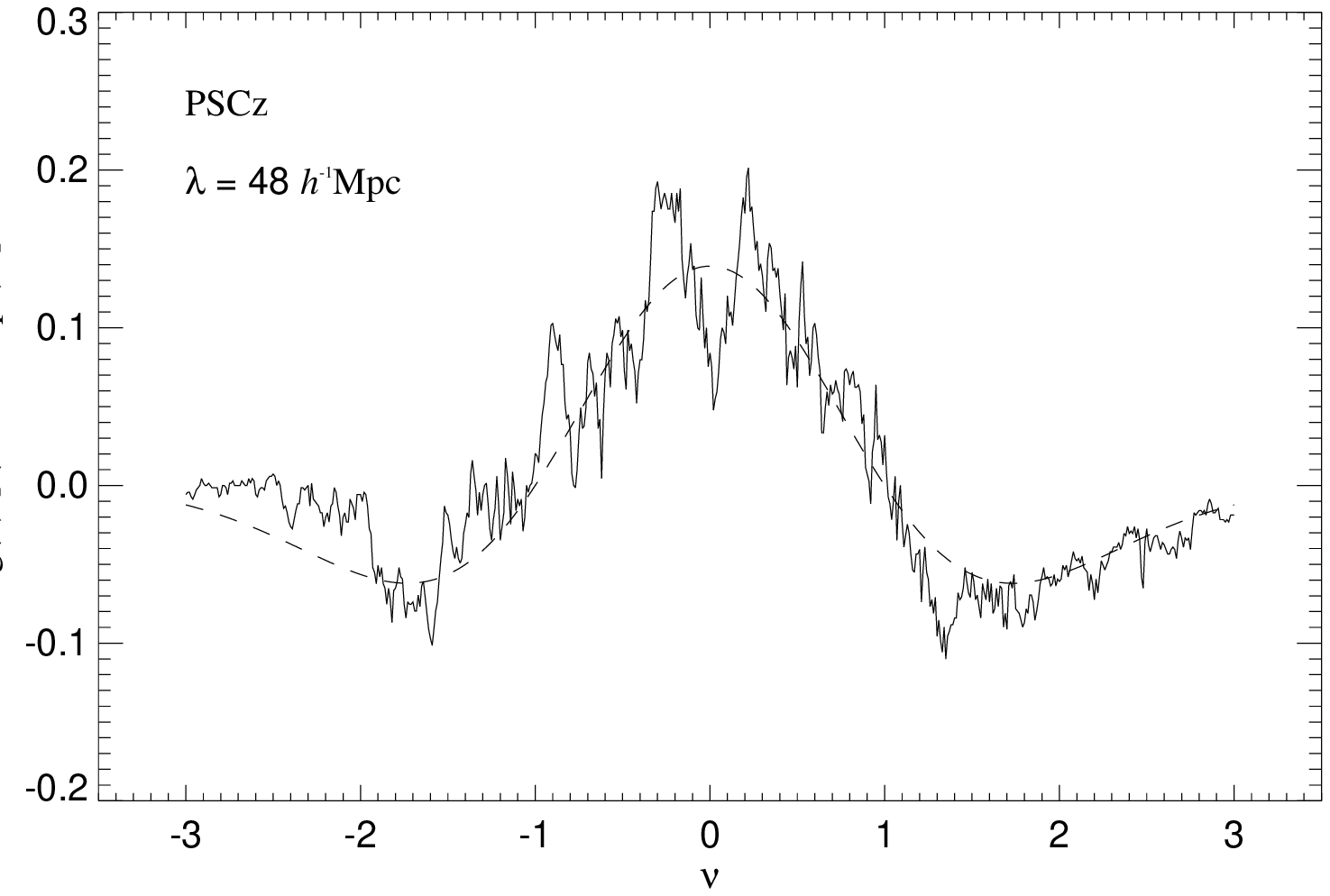}}
\resizebox{8cm}{!}{\includegraphics{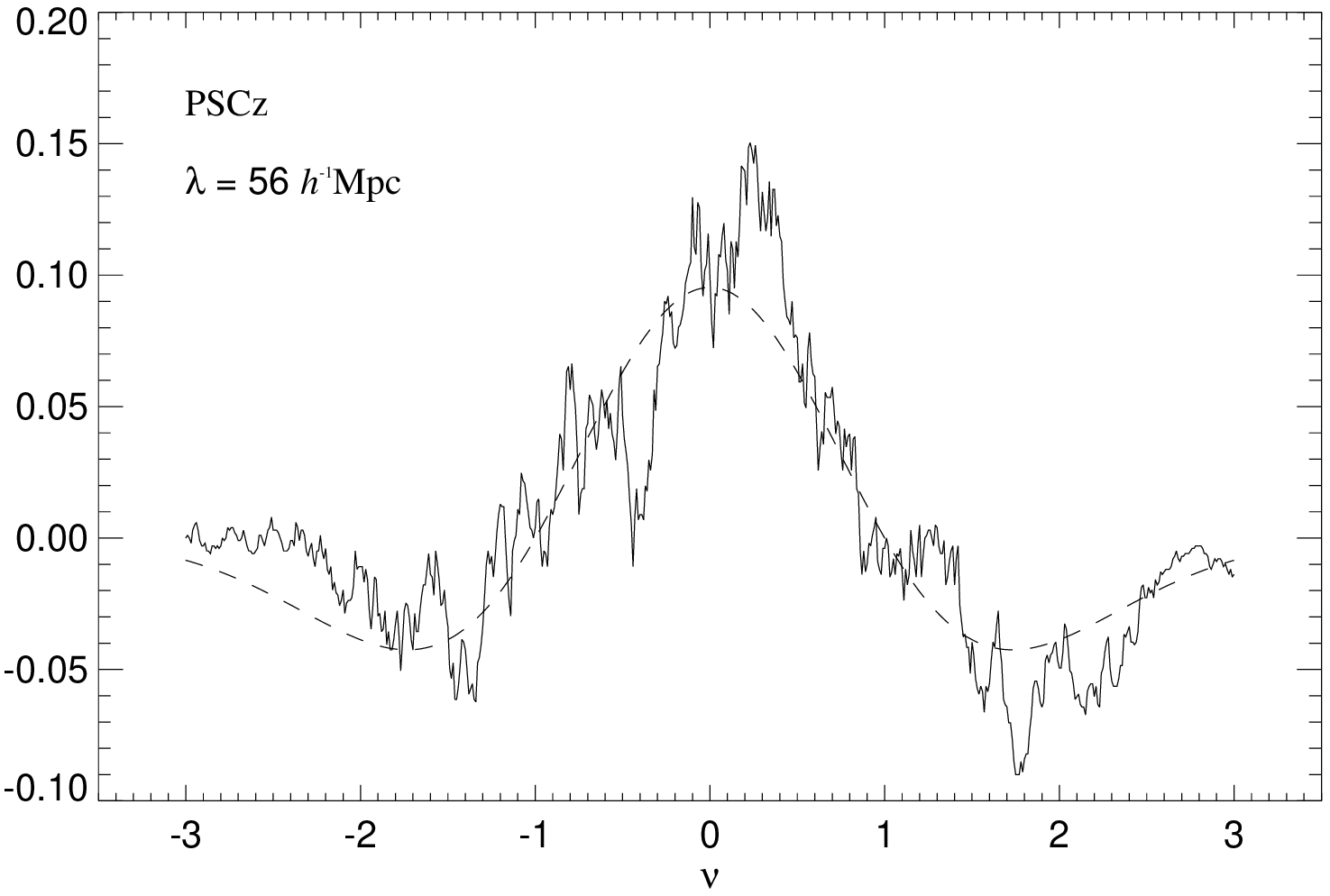}}
\caption{The genus curves of the PSCz redshift survey for selected 
smoothing lengths. In each panel, the solid line represents the raw
PSCz genus, measured at resolution $\Delta\nu=0.01$.
The dotted line gives the best-fit Gaussian curve, 
which also defines the amplitude of the genus curve.\label{genuspscz}
}
\ec
\end{figure*}

Figure \ref{genuspscz} shows the genus curves of PSCz\footnote{The
PSCz genus curves may be
retrieved electronically at http://www.mpa-garching.mpg.de/PUBLICATIONS/DATA/971121\_pscz}  
for smoothing lengths of 
$6$,  $8$, $12$, $16$, $24$, $32$, $48$, and
$56\lu$. The other smoothing scales examined show similar looking
genus curves.
In each panel the solid line shows a 
high resolution curve computed for spacing
$\Delta\nu=0.01$. While it is clear that adjacent points on the genus
curve are highly correlated, the amount of jitter present 
in the curves 
nicely indicates the level of noise contained in the measurements.
Also shown as a dashed line is the best-fit random-phase genus curve,
which we use to infer the measured genus amplitude.

We have chosen not to show the
average genus curve over the set of bootstrap realizations (which
would be a much smoother curve),
 because of the high degree of correlation between
the different points in the curve. 
Furthermore, the bootstrap generated curves are expected to be biased
towards a meatball topology.
However, we use the bootstrap generated curves to estimate  
the statistical uncertainty of individual points on the genus curve.
Similarly, we 
use the bootstrap curves to estimate uncertainties for the
genus meta-statistics. 
For this purpose,
we estimate the 
uncertainty of a particular meta-statistic as the rms-fluctuation of the 
values obtained for the 
set of 15 bootstrap resamplings of the data.

As is evident from the plots in Figure \ref{genuspscz}, the PSCz 
genus curves follow the random-phase expectation rather well. In particular, no obvious deviations
like shifts, broadenings or the like are observed.

\begin{figure}
\bc
\resizebox{8cm}{!}{\includegraphics{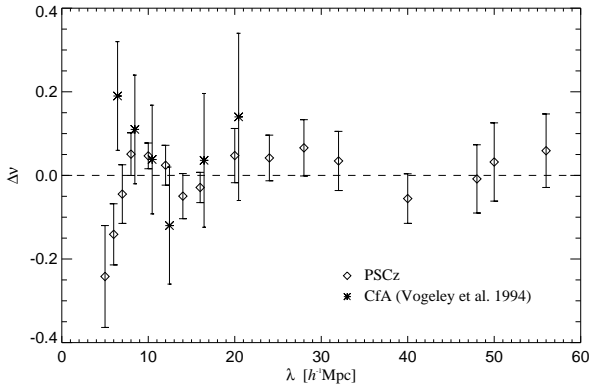}}
\caption{The shift measured for the PSCz genus curves. The error bars
are based on bootstrap resamplings of the data. Also shown are the
results of Vogeley et~al.\ \protect\shortcite{Vo94} for the CfA survey. Here we 
indicate the uncertainty of CfA by the
errors the authors report for mock catalogues extracted from a LCDM
model.
\label{shiftpscz}
}
\ec
\end{figure}

\begin{figure}
\bc
\resizebox{8cm}{!}{\includegraphics{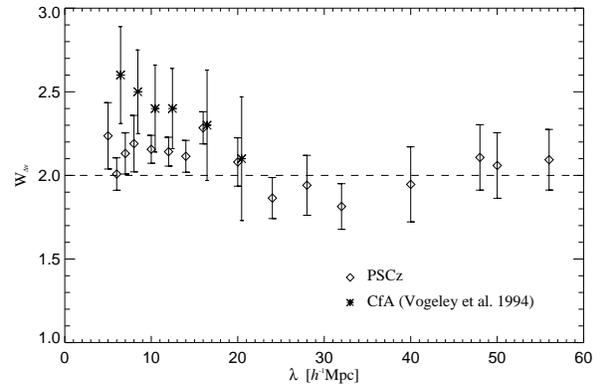}}
\caption{The width of PSCz genus curves, with error estimates derived 
with the bootstrap technique. Vogeley et~al.'s \protect\shortcite{Vo94}
results for the CfA catalogue are shown for comparison.\label{widthpscz}
Here the 
errors are taken to be the uncertainty Vogeley et~al.\ \protect\shortcite{Vo94} report 
for mock catalogues extracted from a LCDM model.
}
\ec
\end{figure}

This is also confirmed by Figures \ref{shiftpscz} and \ref{widthpscz},
which plot the meta-statistics shift and width
for scales ranging from 
$5\lu$ to $56\lu$. The error bars give the rms-fluctuations obtained over $15$
bootstrap resamplings of the data. A comparison is also given with the
results of Vogeley et~al.\ \shortcite{Vo94} for the CfA survey.

Apart from a slight indication of a 
meatball-shift and a broadening of the genus curves at small scales, 
these two meta-statistics fail to show any 
significant departures from Gaussianity.

\begin{figure}
\bc
\resizebox{8cm}{!}{\includegraphics{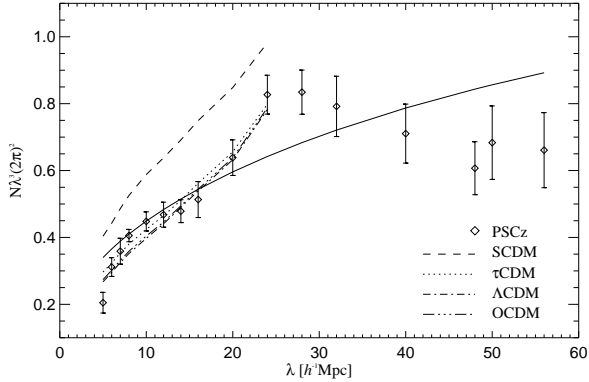}}
\caption{Genus amplitudes for PSCz and for the four CDM models. The
error bars for PSCz represent bootstrap estimates of the sampling
uncertainty.
The solid line gives the expectation based on linear theory for a
power spectrum with shape parameter $\Gamma=0.2$. 
While PSCz matches the $\tau$CDM, $\Lambda$CDM, and OCDM simulations
well, the SCDM model exhibits genus amplitudes that are too large.
\label{amps}
 }
\ec
\end{figure}

\begin{figure}
\bc
\resizebox{8cm}{!}{\includegraphics{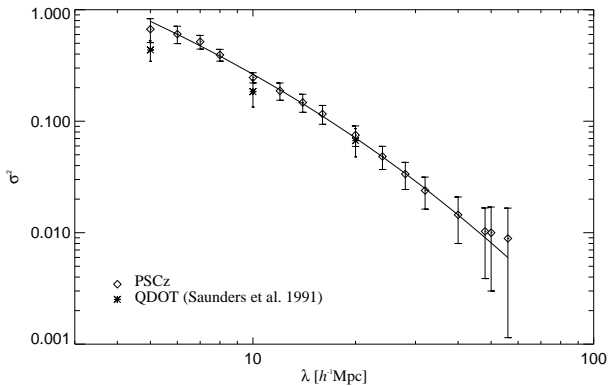}}
\caption{The variance $\sigma^2(\lambda)$ 
found in Gaussian cells for 
the PSCz survey. The $1\sigma$ uncertainties are estimated 
from a set of PSCz-like mock catalogues extracted from a N-body
simulation. 
The solid line represents the best fit of the power 
spectrum (\protect\ref{PowerModels})
to the data, with $\Gamma=0.19 \pm 0.04$ and $\sigma_8=0.84 \pm 0.04$.
For comparison also shown is the result of Saunders
et~al.\ \protect\shortcite{Sa91}
for the QDOT survey (with their error bars).
\label{sigmapscz} }
\ec
\end{figure}

We turn now to the amplitudes of the genus curves themselves. Figure
\ref{amps}
displays our results for PSCz, together with the amplitudes of
variants of CDM models, which we will discuss fully in the next
section. 
Also shown is the expected amplitude 
in linear theory for
the power spectrum (\ref{PowerModels}) with shape parameter $\Gamma=0.2$.
The amplitudes of the simulations are based on the dark matter 
distribution which we identify with the 
galaxy distribution. It should be stressed that the use of volume
weighting makes this 
a relatively weak assumption since we only need regions of 
higher smoothed galaxy density to correspond to regions of higher 
smoothed mass density but not galaxy and mass densities to be 
proportional to one another. The use of volume weighting makes the 
genus curves insensitive to details of bias \cite{Pa91}.

Based on Figure \ref{amps}, 
the standard cold dark matter model in which the galaxies trace 
the mass (SCDM)
is clearly ruled out at high significance level. This can be traced 
back to
the fact that the shape of its power spectrum is in 
strong disagreement with PSCz. The other three N-body models do 
much better in this respect; they match the PSCz amplitudes very well,
because the shape of their power spectrum is in good agreement with PSCz.

It is important to point out the significance of these results. 
Although a counts-in-cell or power spectrum analysis of the PSCz 
redshift survey
will probably lead to the same result for the shape of the power 
spectrum, the route we followed here is radically different. The
genus depends on the spatial coherence of structures rather than
on the
rms amplitude of fluctuations. 
Moreover, it depends only on the rank order of density values, 
not on 
the values themselves. In the presence of unusual forms of
biased 
galaxy formation, non-Gaussian initial conditions, or observational
errors, 
there is no guarantee {\em a priori} that the determination of the
shape 
of $P(k)$ via the genus amplitude will be consistent with that infered
from a
direct measurement of $P(k)$. The fact that the same conclusion is 
reached by a very different analysis of the data lends considerable, 
largely independent support to the $P(k)$ result.

Note that, in Figure \ref{amps}, we have plotted the
amplitude $N$ of PSCz without a shot-noise correction, i.e. these
values are likely to be biased slightly {\it high}.

\subsection{Variance and amplitude drop}

Figure \ref{sigmapscz} shows our results for the variance of the
Gaussian smoothed PSCz density fields, 
obtained with the unbiased estimator described in
Appendix \ref{appvar}. The error estimates are based on an ensemble
of PSCz-like mock surveys drawn from a N-body simulation. 
Note that we find somewhat more power on small scales than was found
for QDOT \cite{Sa91}.

A fit of the linear CDM power spectrum (\ref{PowerModels}) to the
measured variances results in a shape parameter $\Gamma=0.19\pm
0.04$ and a normalization $\sigma_8=0.84\pm 0.04$, here 
expressed in terms of the rms-fluctuations in
top-hat spheres of radius $8\lu$. 
Note that these
values refer to redshift-space only; no correction for redshift space
distortions has been attempted.
The quoted
uncertainties are estimated by approximating 
the measurements for different $\lambda$ as being independent. 

As outlined in Section \ref{drop} we also compute estimates for the
logarithmic slope of $\sigma^2(\lambda)$ and the approximate
shot-noise correction factor $A_{\rm shot}$.
Combining these measurements 
allows us to estimate the amplitude drop of PSCz as $R=N^*/N_{\rm rp}=N/(A_{\rm
shot}N_{\rm rp})$. 

The uncertainty of the measured values for $R$ is difficult to estimate,
since we deal with a ratio of correlated quantities. We tentatively
assign the relative error of the amplitude measurement as error of
$R$. 
Our
experiments with mock catalogues suggest that this gives approximately
the right size of uncertainty.

\begin{figure}
\bc
\resizebox{8cm}{!}{\includegraphics{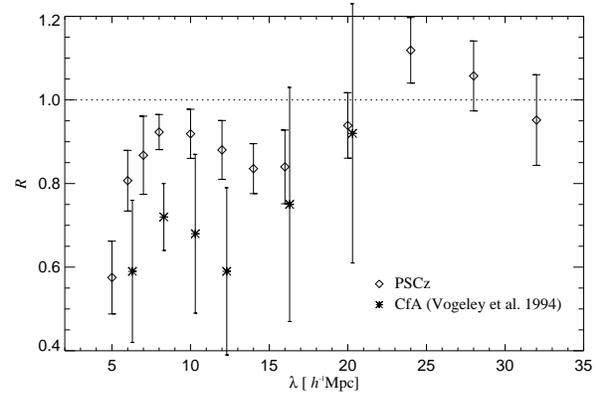}}
\caption{The amplitude drop $R$ measured for PSCz. Also shown are the results
obtained by Vogeley et~al.\ \protect\shortcite{Vo94} for the CfA survey.
Here the 
errors are taken to be the uncertainty Vogeley et~al.\ \protect\shortcite{Vo94} report 
for mock catalogues extracted from a LCDM model.
\label{droppscz}
}
\ec
\end{figure}

In Figure \ref{droppscz} 
we plot our results for the amplitude drop of PSCz.
Although some amount of phase correlation seems to be  
present at small scales, 
we find that PSCz does not exhibit strong phase correlations on scales
above $10\lu$, in contrast to the findings of Vogeley
et~al.\ \shortcite{Vo94} for the CfA survey.
This means that the PSCz density
field is consistent with random-phase initial conditions. In the next
section we will further analyse this amplitude drop and compare it to
the results of 
N-body simulations.

\section{N-body simulations of CDM Models}

        \label{nbody}

\begin{figure*}
\bc
\resizebox{8cm}{!}{\includegraphics{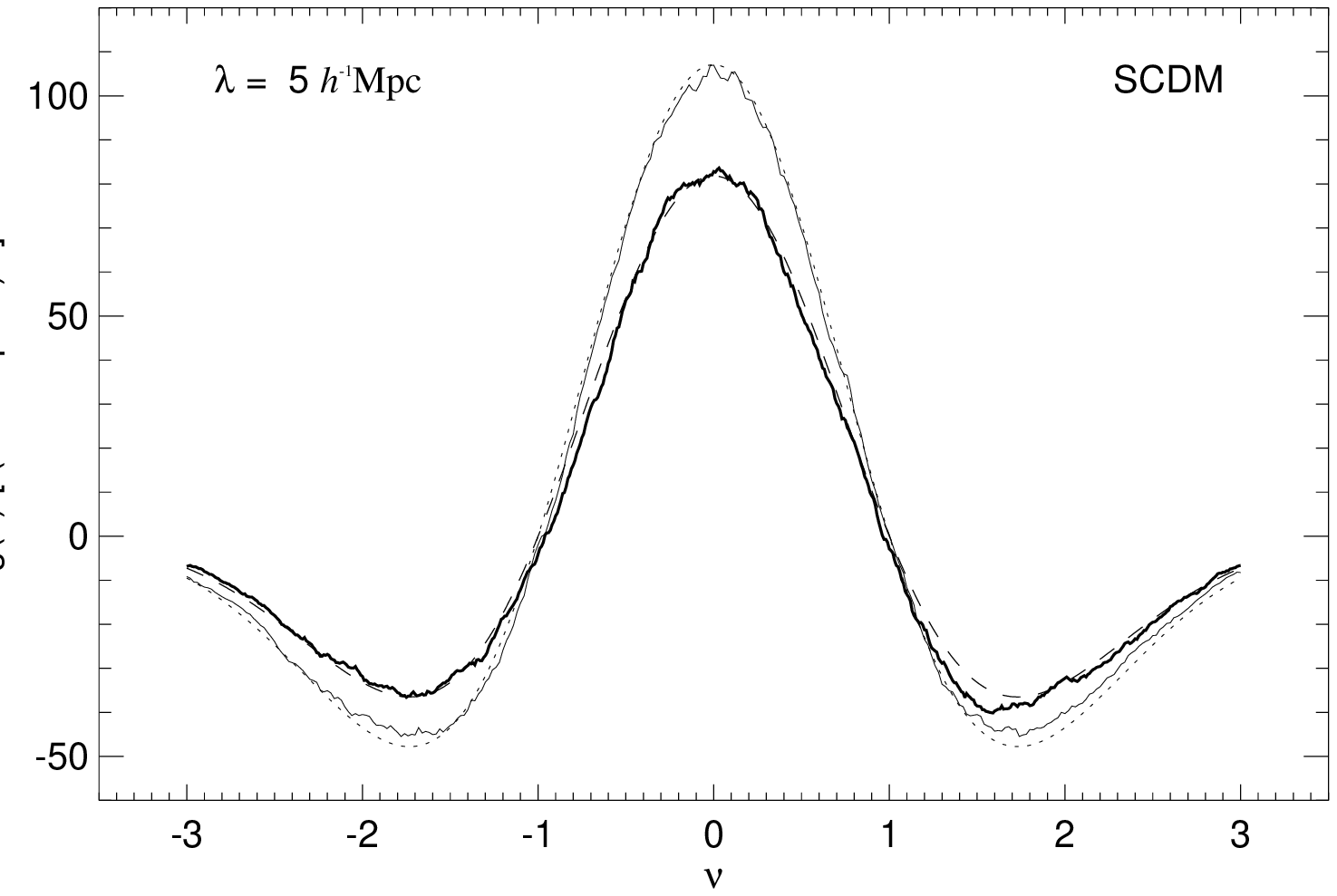}}
\resizebox{8cm}{!}{\includegraphics{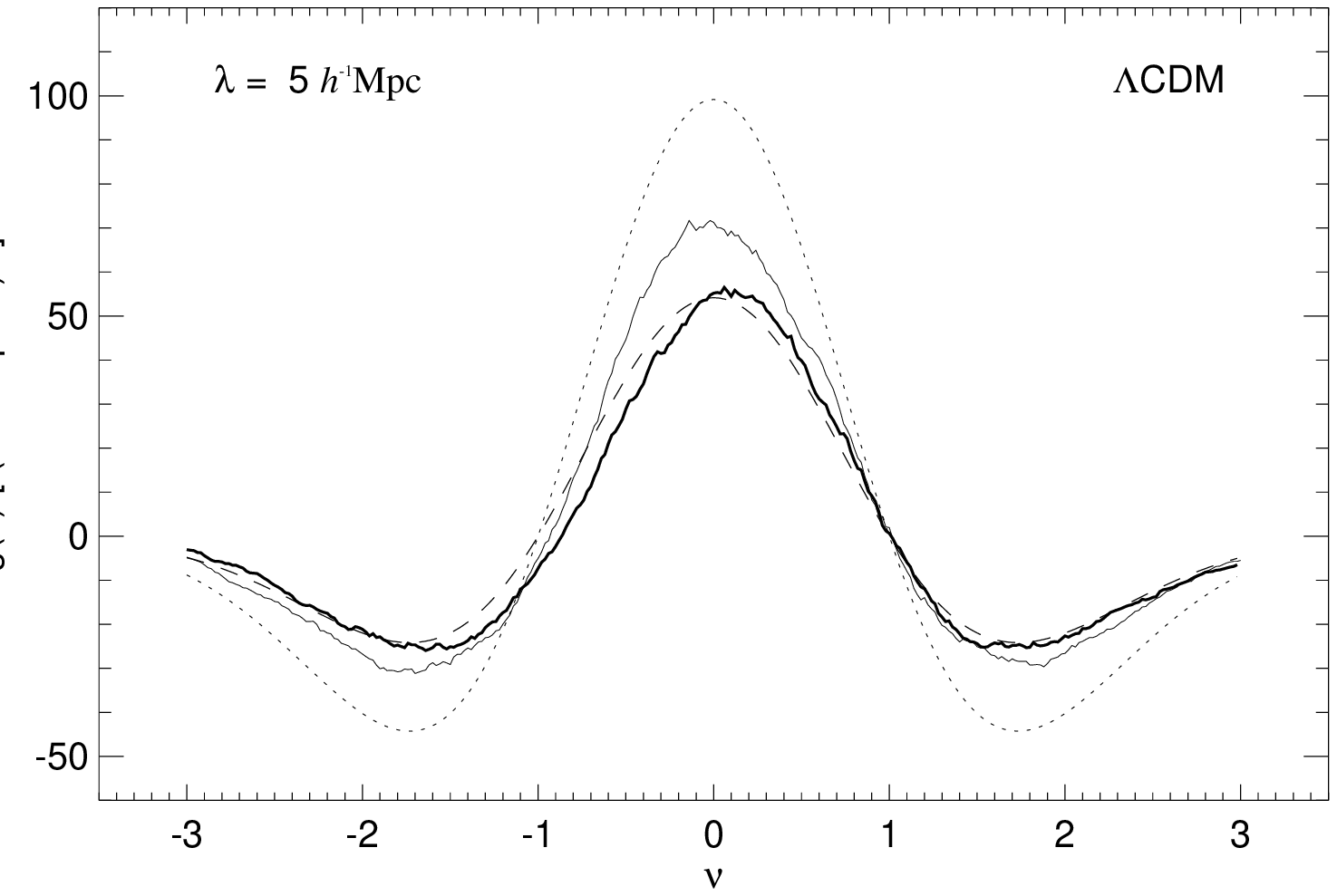}}
\resizebox{8cm}{!}{\includegraphics{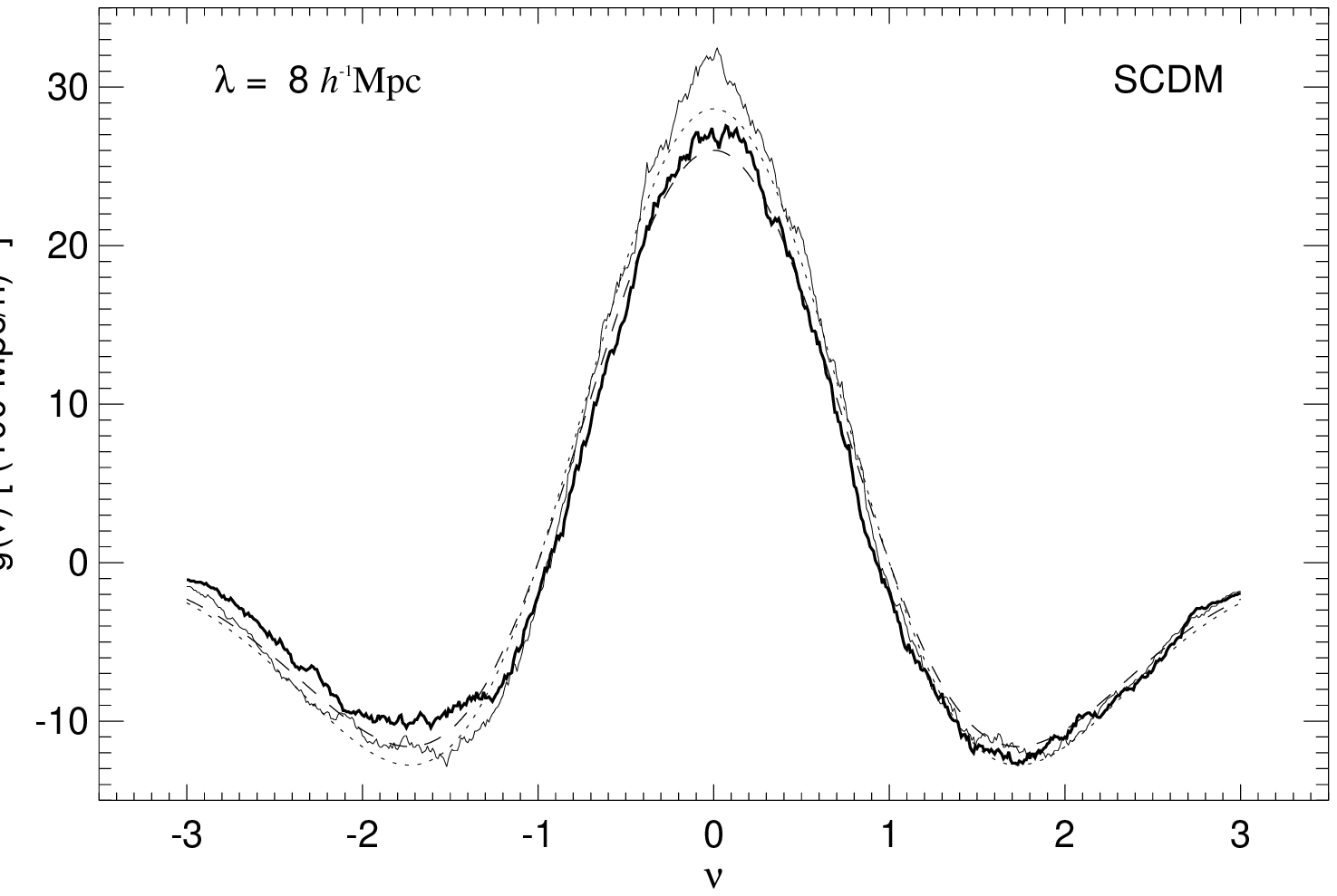}}
\resizebox{8cm}{!}{\includegraphics{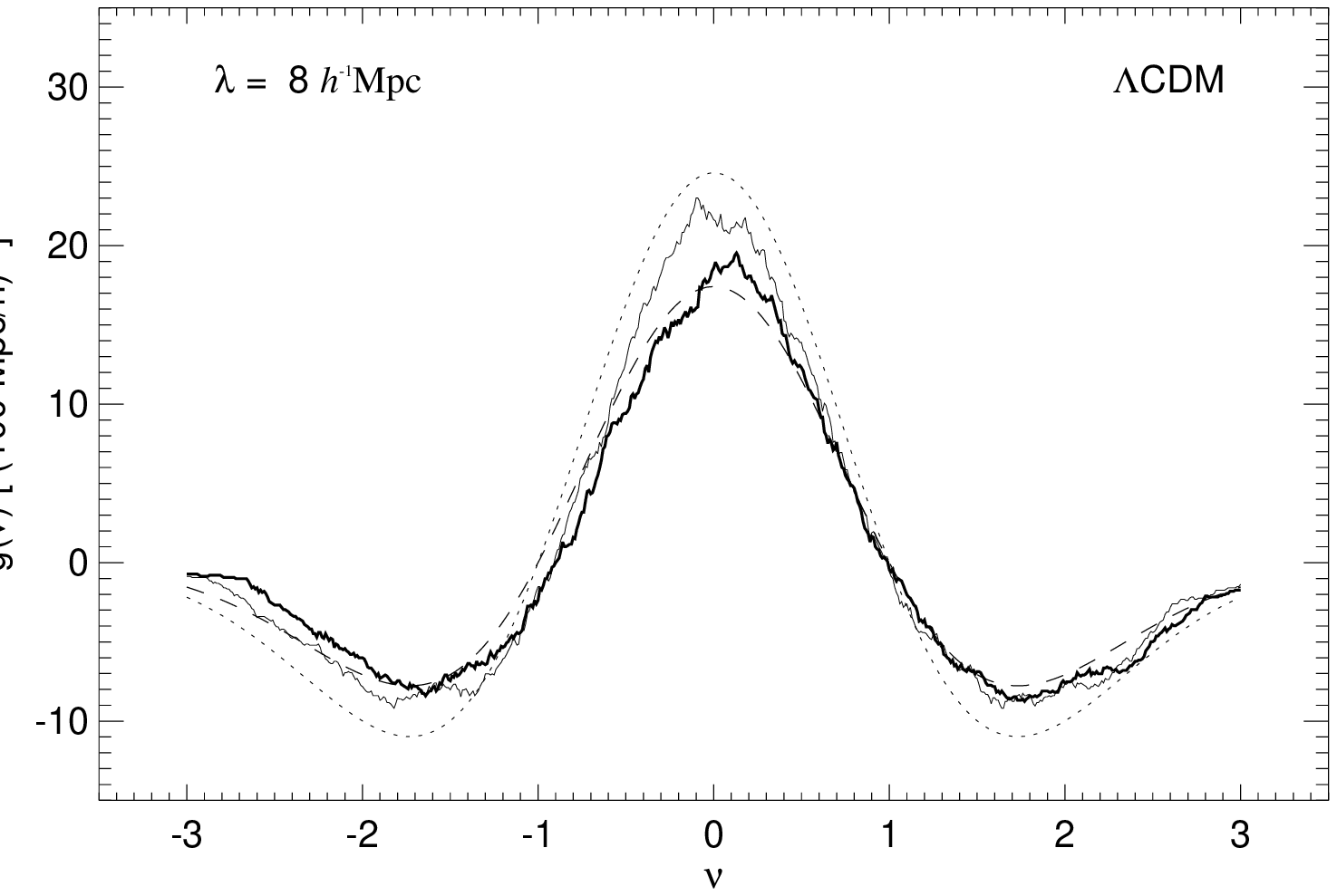}}
\resizebox{8cm}{!}{\includegraphics{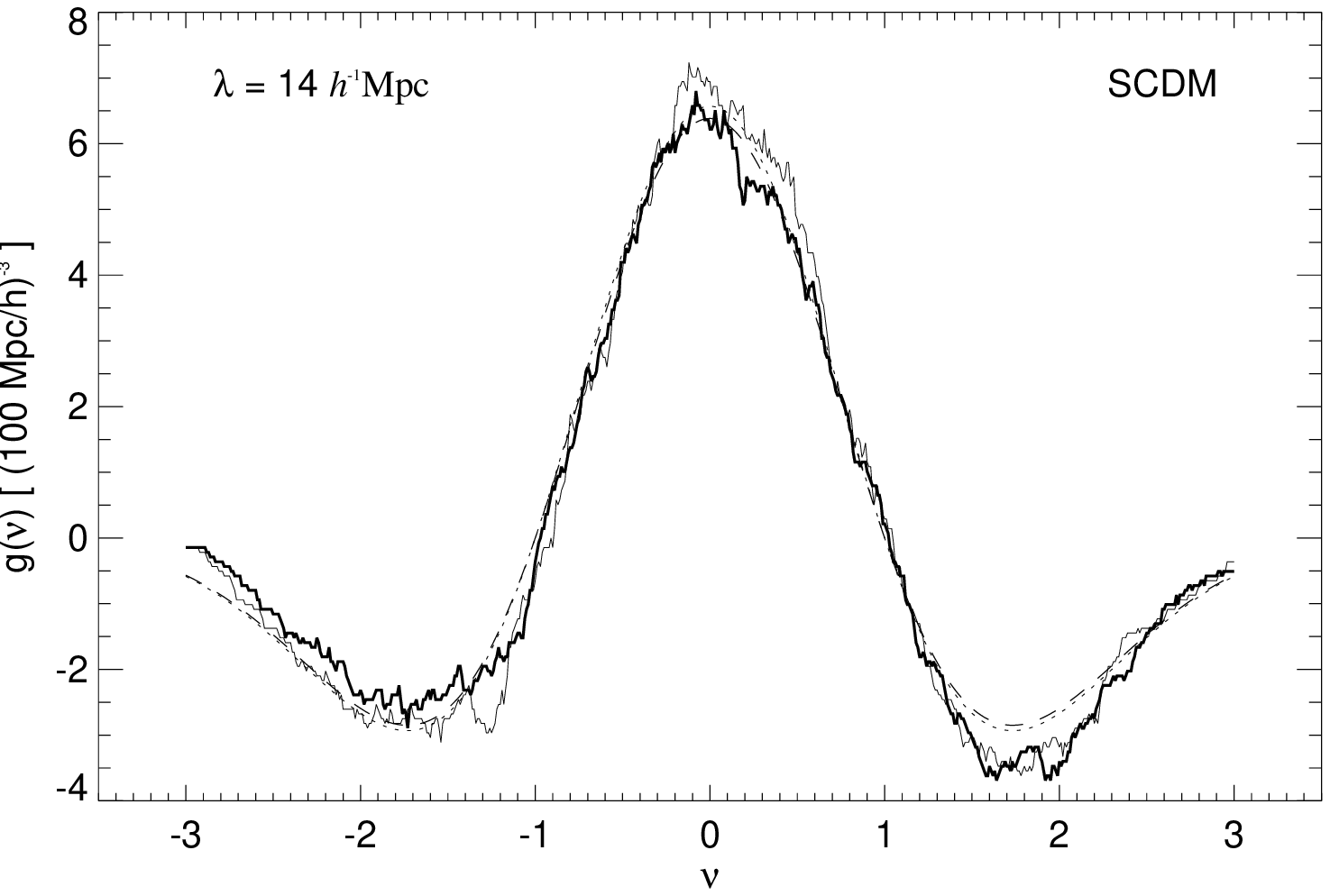}}
\resizebox{8cm}{!}{\includegraphics{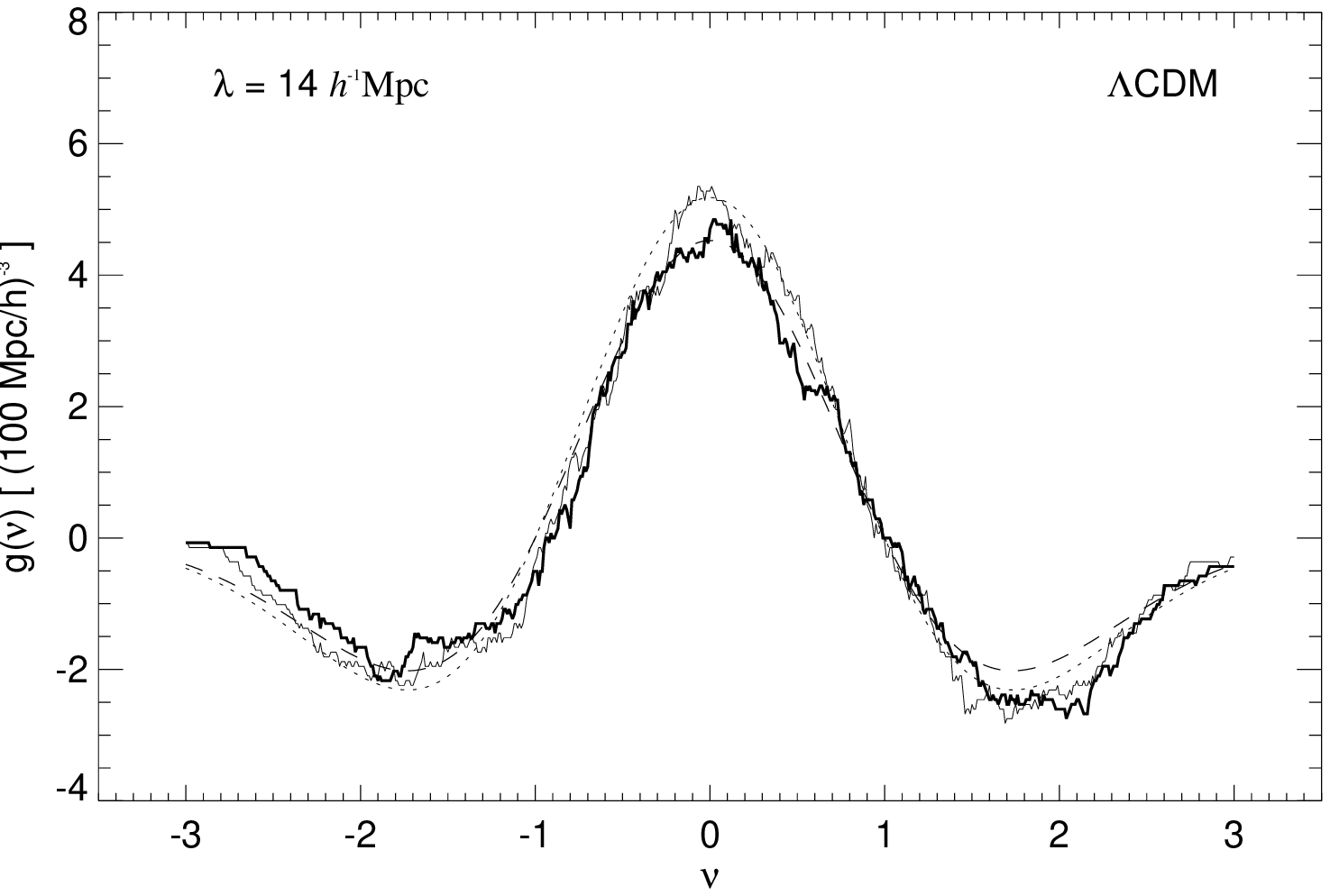}}
\resizebox{8cm}{!}{\includegraphics{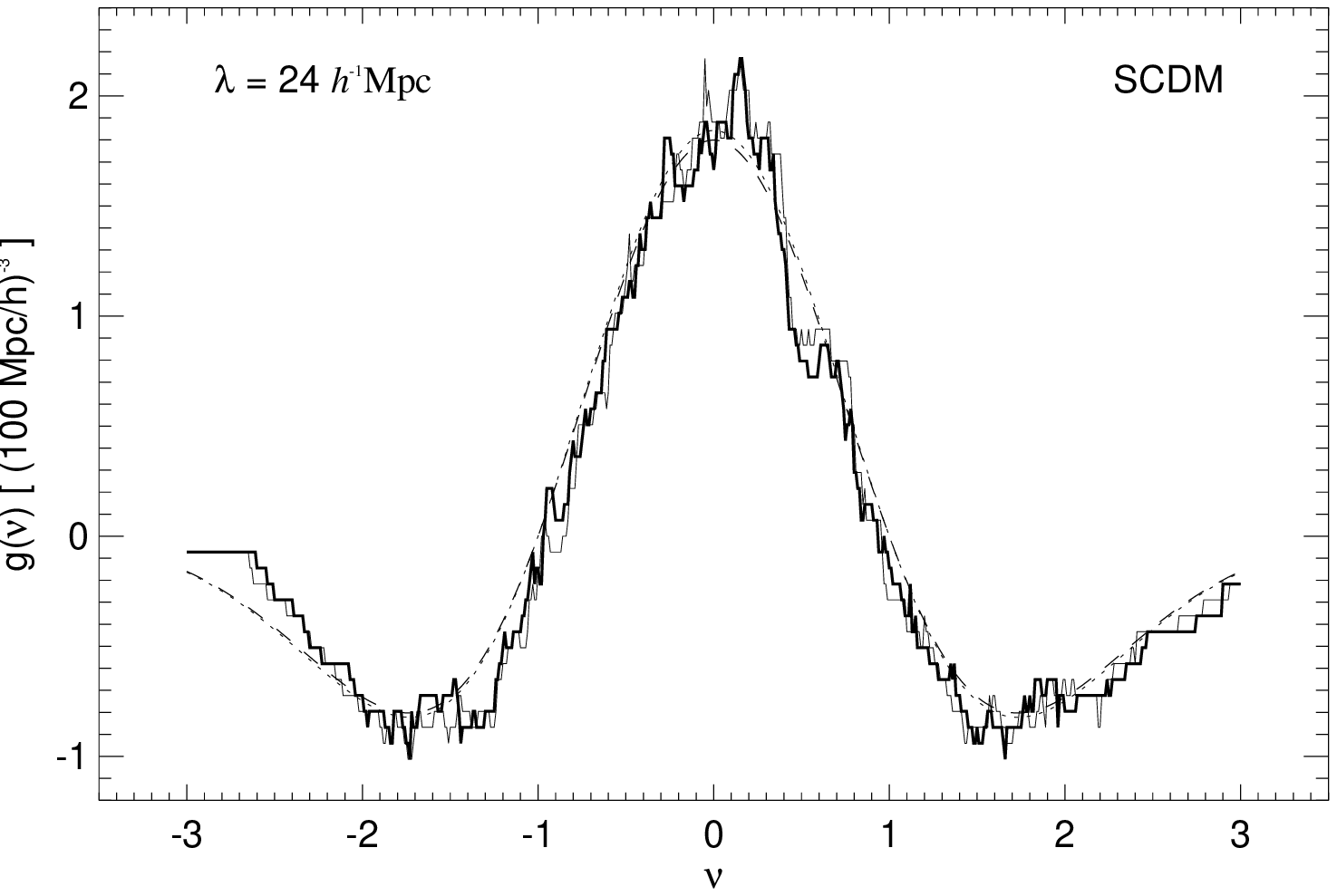}}
\resizebox{8cm}{!}{\includegraphics{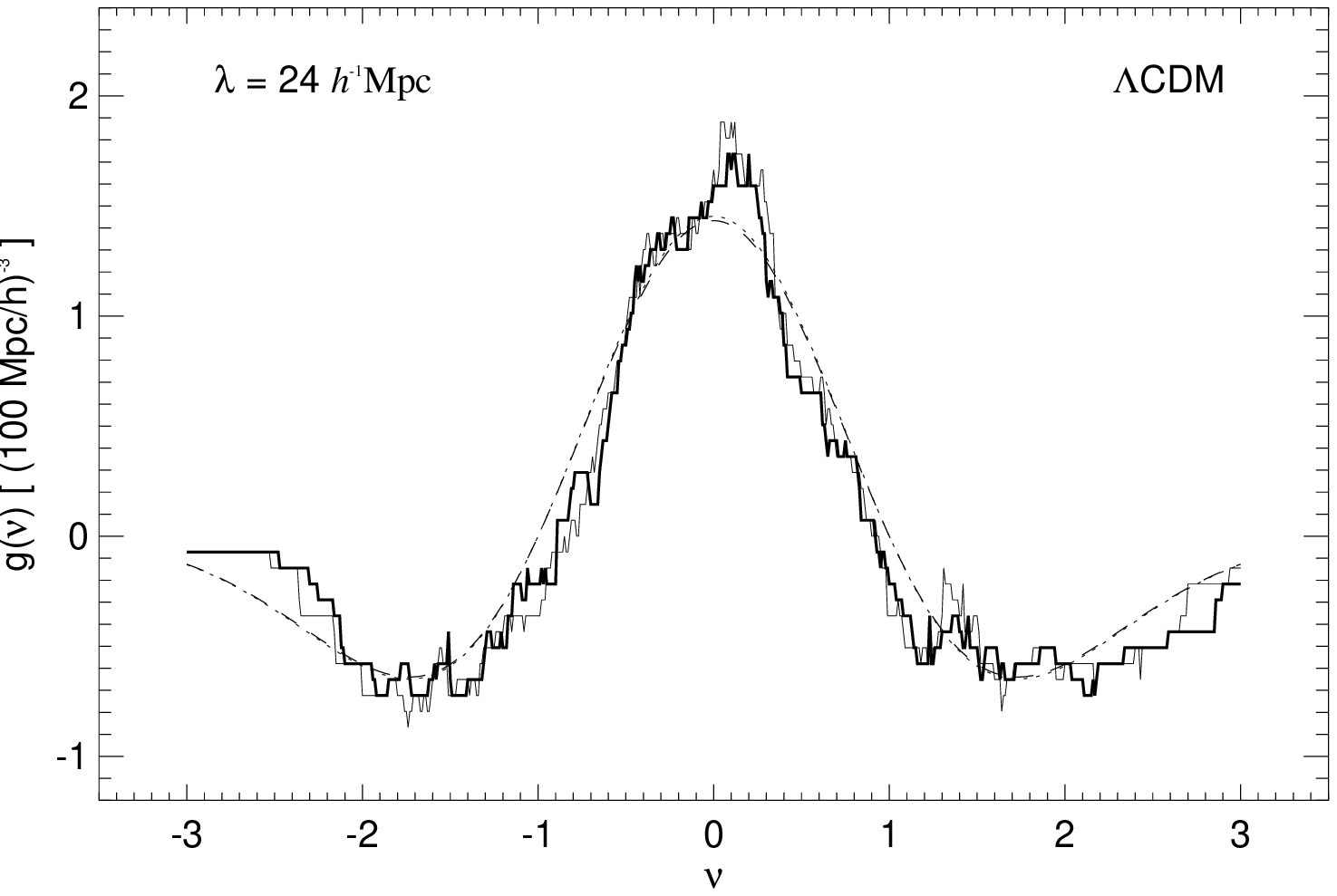}}
\caption
{Genus curves of the SCDM (left column) and $\Lambda$CDM (right
column) simulations 
at selected smoothing lengths. In each panel we show 
the actual genus curve of the evolved simulation as 
heavy solid line, and we
give the best-fit random-phase genus curve
as dashed line.
The dotted line is the genus 
of the corresponding {\it Gaussianized} field, while the thin solid
line
shows the genus curve of the initial conditions.
\label{GL}\label{GS}}
\ec
\end{figure*}

\subsection{The simulations}

We study four variants of the cold dark matter cosmogony. 
These models were kindly provided by the Virgo collaboration
\cite{Je97,Col97} 
who have recently started an ambitious project to model the formation of structure
in the Universe making use of the largest available supercomputers in 
Europe. The simulations have been performed with an  AP$^{3}$M-SPH code 
named {\small HYDRA} \cite{Cou95}.

All four models we examine 
contain only cold dark matter in periodic boxes of size $239.5\lu$.
The basic parameters of the individual runs are given in Table
\ref{modelparameters}. 
Here the shape parameter $\Gamma$ refers
to the linear theory power spectrum
\be
\label{PowerModels}
P(k)=\frac{Bk}{\left(1+[ak+(bk)^{3/2}+(ck)^2]^\nu\right)^{2/\nu}} ,
\ee
where $a=6.4\,\Gamma^{-1} \lu$, $b=3.0\,\Gamma^{-1} \lu$, 
$c=1.7\,\Gamma^{-1} \lu$ and $\nu=1.13$ \cite{Bo84,Ef92}. 
The models are normalized so as to match the observed abundance of 
rich clusters of galaxies \cite{Ek96}.

The standard cold dark 
matter model (SCDM) has the power spectrum ($\ref{PowerModels}$) 
with a shape parameter $\Gamma=0.5$ in the linear regime, 
as predicted from a CDM inflationary scenario with a 
primordial scale-invariant Harrison-Zel'dovich spectrum. 
The $\tau$CDM model is a variant of CDM with more 
power on large scales, as suggested by numerous recent 
observations \cite{Sa91,Pe94,Ol96}.
The same shape of the power spectrum may also be 
obtained in low-density universes.
We consider two models of this class;
one 
open universe (OCDM), and one
with a cosmological constant ($\Lambda$CDM) that retains a flat
background geometry.

\begin{table}
\bc
\caption{Parameters of the examined CDM models. The simulations have
been done by the 
Virgo collaboration.\label{modelparameters}
}
\begin{tabular}{l|c|c|c|c|}
\multicolumn{1}{l|}{ }& SCDM & $\tau$CDM & $\Lambda$CDM & OCDM\vspace{0.1cm}\\ 
\multicolumn{1}{l|}{Number of particles } & $256^{3}$ & $256^{3}$ & $256^{3}$ & $200^{3}$ \\
\multicolumn{1}{l|}{Box size$[\lu]$ } & $239.5$ & $239.5$ & $239.5$ & $239.5$ \\
$z_{start}$ & $50$ & $50$ & $30$ & $119$ \\
$\Omega_{0}$ & $1.0$ & $1.0$ & $0.3$ & $0.3$ \\
$\Omega_{\Lambda}$ & $0.0$ & $0.0$ & $0.7$ & $0.0$ \\
 \multicolumn{1}{l|}{Hubble constant $h$} & $0.5$ & $0.5$ & $0.7$ &$0.7$\\
$\Gamma$ & 0.5 & 0.21 & 0.21 & 0.21 \\
$\sigma_{8}$ & $0.60$ & $0.60$ & $0.90$ & $0.85$ \\

\end{tabular}
\ec
\end{table}

The simulations contain typically more 
than 16.7 million particles, thus providing 
excellent data which allow very accurate measurements of the genus 
statistic, 
especially on small scales, where the number of resolution elements 
is high and the genus curves are hardly influenced by cosmic
variance. 
This is, however, not true for scales $\geq 20 \lu$. 
The number of resolution elements in the simulations drops from 
$19861$ for $\lambda=5\lu$ to $180$ for $\lambda=24\lu$. Hence 
the survey volume is quite limited even for such large simulations. 
In fact, beyond $20 \lu$
the volume probed by PSCz becomes larger than that of 
the simulations.  We therefore restrict our genus analysis of the
models 
to the ten smoothing scales between 
$5 \lu$ and $24 \lu$ used in the analysis of PSCz.

In order to construct smoothed density fields for the simulations 
we bin the particles on a $128^3$ grid by CIC assignment and 
subsequently smooth with a 
Gaussian kernel. We also compute 
{\it Gaussianized} representations
of the simulations by taking the Fourier transforms of the original
density fields, randomizing the phases in Fourier space subject to the
reality constraint ${\delta_{k}}^*=\delta_{-k}$ and transforming 
back to real space. 
These fields are then smoothed in the same way as the original ones, and 
the genus curves are calculated as before. 
However, due to the limited simulation volume
there is some variance introduced in the resulting genus curves
from the variety of different possible random-phase realizations. In
order to reduce this effect to a minimum we 
compute 10 Gaussianized fields and take the average genus curve of
those as the Gaussianized
genus curves. The amplitude of the latter is used to estimate the
amplitude drop.

\subsection{Results}

\begin{figure}
\bc
\resizebox{8cm}{!}{\includegraphics{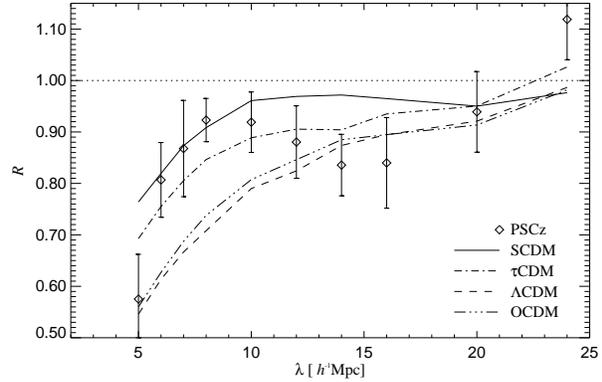}}
\caption
{The amplitude drop measured for the four N-body simulations as a
function of the smoothing scale. Also shown are the values estimated
for the PSCz survey (and error bars).
\label{modeldrops1}
}
\ec
\end{figure}

\begin{figure}
\bc
\resizebox{8cm}{!}{\includegraphics{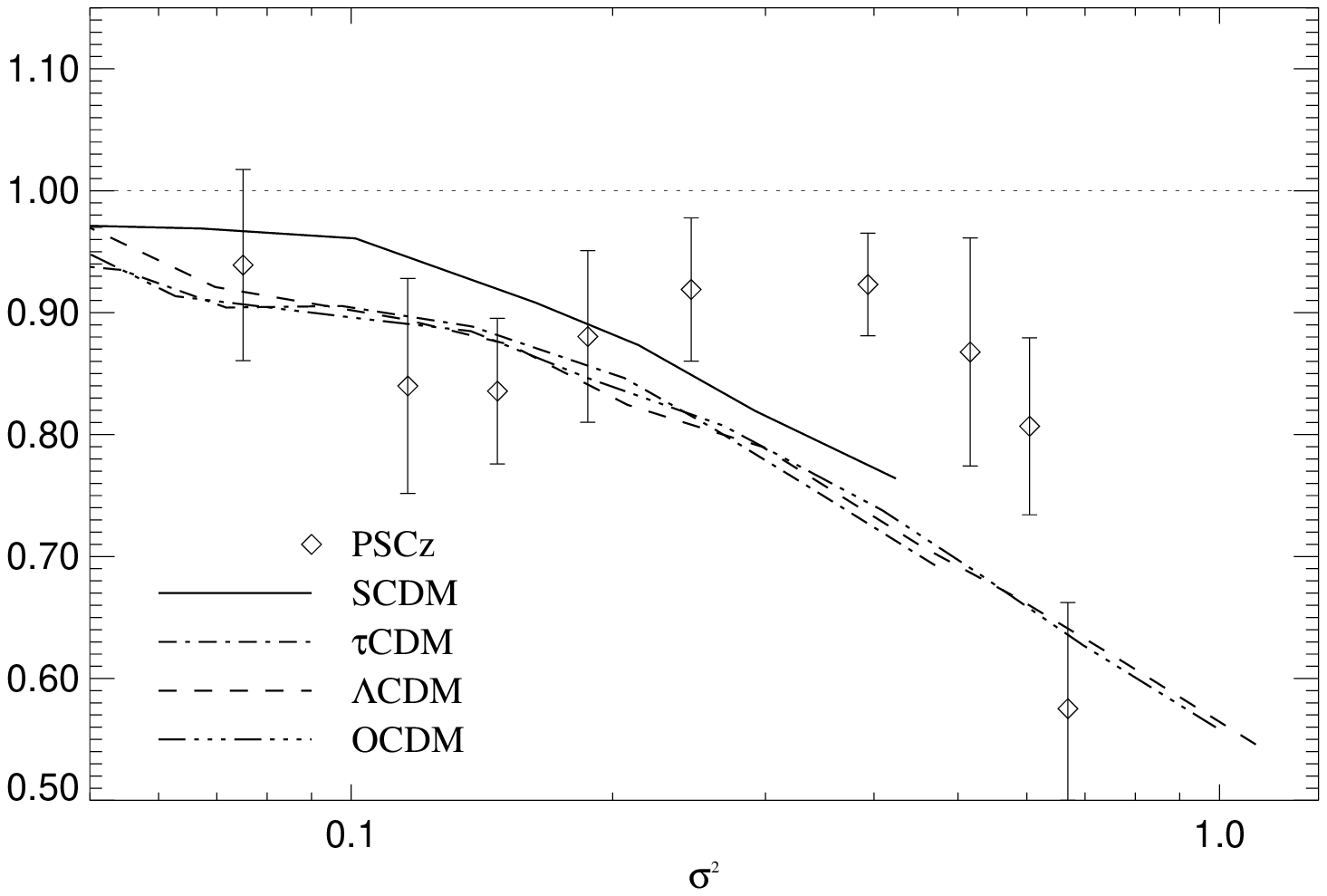}}
\caption
{The amplitude drop revisited. Here we plot the data of Figure
\protect\ref{modeldrops1}
again, however this time as a function of the variance of the smoothed
density fields. The good match of the three $\Gamma=0.21$ models in
this representation suggests that the differences that show up in
Figure \protect\ref{modeldrops1} between these models 
are only due to their slightly 
different normalizations. The SCDM model ($\Gamma=0.5$),
however, continues to show a considerably smaller amplitude drop.
\protect\label{modeldrops2}
}
\ec
\end{figure}

Figure~\ref{GS} shows
representative genus curves obtained for the SCDM and $\Lambda$CDM models.
In each panel the 
heavy solid line
gives the genus curve for the density field of the evolved  
simulation at the indicated smoothing scale.  
A fit of these data to the generic random-phase genus curve is
shown as a dashed line. 
The dotted line shows the genus curve for
the  corresponding {\it Gaussianized} field, 
and the thin solid line marks the genus of
the initial conditions of the simulations. 
For the sake of brevity we don't show further genus
curves for the $\tau$CDM and OCDM models. They look qualitatively very
similar to those in Figure~\ref{GS}.

Note that there is no sampling noise in these density
fields; the residual tremble that is present in the N-body genus
curves is caused by the finite number of resolution elements,
i.e.\ ultimately by cosmic variance.

It is apparent that all
the genus curves follow the Gaussian w-shape very well. Apart from the
small bubble-shift of the $\Lambda$-model at the smallest smoothing
scales
the only meta-statistic able to measure a significant 
deviation from Gaussianity is the amplitude. 
It is interesting to note that 
the Gaussianized 
genus curves of the SCDM model
remain
close to the ones of the initial linear density field
at all smoothing scales.
In contrast, the $\Lambda$CDM model shows 
signs of non-linear evolution 
of the power spectrum at 
$\lambda=5\lu$.
Its effect has been to increase
the effective slope of the power spectrum
\cite{Je97} 
on the relevant scales, and thereby also to increase the
genus amplitude of the Gaussianized fields. 
Higher-order correlations,
however, `keep the genus in place', and decrease the
amplitude at small $\lambda$ 
below the linear theory expectation.

We defined the amplitude drop as the ratio of the genus amplitudes of
the evolved density field and its Gaussianized counterpart. Hence, this
quantity is a quantitative measure of higher-order correlations.
In all of the 
four models
an amplitude drop is 
present 
at small scales, being largest for the 
$\Lambda$CDM model and smallest for the SCDM model.

In Figure \ref{modeldrops1} we plot this amplitude drop against the
smoothing scale for the four simulations.  
As expected, the amplitude drop becomes smaller with increasing 
smoothing length. This is consistent with the picture that all
non-linear features are sufficiently smoothed out in this regime.
It is interesting however, that the models exhibit a varying strength 
of the amplitude drop. For the $\Lambda$CDM and OCDM models,
phase correlations up to the scale of  $24 \lu$ are detected, 
whereas the SCDM 
becomes very close to Gaussian at about $10 \lu$, and the $\tau$CDM takes an
intermediate position.

Figure  \ref{modeldrops2} shows that part of this difference is caused
by the different normalizations of the models. Here we have plotted
the amplitude drop against the variance $\sigma^2(\lambda)$ of the
density fields. The good match between the models with shape parameter
$\Gamma=0.21$ suggests that the strength of the amplitude drop at
fixed $\sigma$
is primarily governed by the shape of the power spectrum.

In Figures \ref{modeldrops1} and \ref{modeldrops2} we have also
plotted the measured amplitude drops for the PSCz
survey. It is encouraging that the measurements for PSCz exhibit about
the right strength of phase correlations like the one expected in CDM
scenarios. In principle, one could have hoped that this measurement
might be able to strongly favour one of the CDM variants. However, at
this point the discriminative power of the genus test is not strong
enough for this. The $\Lambda$CDM, $\tau$CDM, and OCDM fit the PSCz
almost equally well. Only Figure \ref{modeldrops1} suggests a slight
preference for $\tau$CDM, essentially favouring its
normalization.  
Given the considerable uncertainties in the biasing
issue, this is not more than circumstantial evidence for this model.
Note that in contrast to Figure \ref{modeldrops1}
the horizontal positions of the PSCz points in 
Figure \ref{modeldrops2} depend
on bias; a positive bias factor would reduce the mismatch between PSCz
and the models at large $\sigma^2$.

It should be noted, however, that we are comparing 
observations with {\em perfect} simulation data. 
Mock catalogues should be used to
accurately assess systematic effects
from shot noise, redshift space distortions, selection function, 
and so forth, so that one can 
establish firm conclusions about the 
viability of models. 
Protogeros \& Weinberg \shortcite{Pr97} have pioneered this approach,
and it will also be applied 
in a forthcoming paper by Springel et~al.\ \shortcite{Sp97b}.

\section{Conclusions}

\label{conclusion}

We have analysed the topology of a new large redshift survey, the
PSCz, 
by means
of genus statistics. The genus test provides not only information
about the shape of the power spectrum but is also sensitive to higher
order correlations of the density field. The PSCz survey is well
suited for the genus test, due to its large survey volume and its near
full sky coverage. In particular, 
it allows a topological analysis with larger statistical significance 
than any previously
examined sample.

The genus curves of PSCz are featureless at smoothing scales ranging
from $5\lu$ to $56\lu$, i.e.\ they exhibit the w-shape that is characteristic 
of random-phase
density fields. In particular, we find no clear evidence for shifts or
broadenings of the genus curves. Evidence for significant non-Gaussian
signatures on large scales found previously in 
Vogeley et~al.\ \shortcite{Vo94} are not confirmed by PSCz.

When the genus amplitude is examined, we 
find that the PSCz density field is consistent with a CDM
power spectrum with shape parameter $\Gamma\approx 0.2$ 
and inconsistent with the
SCDM model. These results seem 
consistent with previous analysis of the shape of the power 
spectrum based on topology \cite{Mo92,Vo94,Pr97}. We expect that 
an ordinary power
spectrum analysis of PSCz will give a similar result. 
Again, we 
want to stress that these two methods of estimating 
the shape of the power spectrum are largely independent.

We have demonstrated that the genus curves of 
CDM models retain the w-shape of the random
phase genus curve even at small smoothing lengths where
non-linear gravitational clustering has already generated significant
skewness of the one-point PDF.
However, the non-linear
evolution manifests itself in a depressed genus amplitude compared to
the expectation based on the power spectrum alone. This amplitude drop 
directly quantifies the amount of higher order correlations present in
the density field. When plotted against the variance of the smoothed
density field, we find that the amplitude drop depends on the shape of
the power spectrum alone.

The measured amplitude drop for PSCz is consistent with the CDM models,
i.e.\ the non-detection of strong phase correlations in PSCz 
for smoothing scales above $10\lu$ supports the hypothesis that
structure grew from random-phase initial conditions. 
Alternative models for structure formation that provide a 
non-Gaussian seed field (for example 
by means of cosmic strings or explosions)
are
expected to exhibit stronger higher-order correlations 
in their density fields. Future analysis of the genus 
of such models should be used to test whether these correlations
indeed show up as a strong amplitude drop or a distortion of the shape
of the genus curve. In this case, our results for PSCz 
can place strong limits on the
viability of such models.

\section{Acknowledgements}

We thank the referee, David Weinberg, for many helpful comments on
this paper.
AC acknowledges the support of JNICT (Portugal). CSF acknowledges a 
PPARC Senior Research Fellowship. We are grateful to the Virgo 
consortium (J. Colberg, H. Couchman, G. Efstathiou, C. S. Frenk, 
A. Jenkins, A. Nelson, J. Peacock, F. Pearce, P. Thomas and
S. D. M. White)
for providing their N-body simulations in advance of publication, and
to J\"{o}rg Colberg for many interesting discussions.

\bibliography{paper}

\appendix

\section{Clustering amplitude determination}

\label{appvar}

In this appendix we outline our method to calculate the 
variance $\sigma^2(\lambda)$ of the smoothed PSCz density field.
This measurement is used to estimate the shot noise influence on
the genus amplitude and to obtain an estimate of  
the genus amplitude 
of the {\it Gaussianized} version of the PSCz density field. 
Our approach 
is that of Springel \& White \shortcite{Sp97a} which
is a slightly modified version of the method proposed 
by Saunders et al.\ \shortcite{Sa91}.

We start with an unsmoothed density field derived by binning the
galaxies
on a fine mesh.
We assume that the probability of 
finding a galaxy in a 
voxel $i$ of volume $\delta V_{i}$ 
is given by $p\propto\rho_{i}S_{i}\delta V_{i}$ 
where $\rho_{i}$ is the density and $S_{i}$ 
the value of the selection function 
in the cell.
For convenience, we assume $\rho=1$ in the following. 
Then the 
probability density function that describes 
the distribution of the counts in cell $i$ is given by 
\be
p_{i}(m)= \sum_{N=0}^{\infty}\frac{\overline{m_{i}}^{N}}{N!}
\,{\rm e}^{-\overline{m_{i}}}\delta^{\rm (D)}(m-N).
\ee
Here $\overline{m_{i}}=\rho_{i}S_{i}$ denotes the 
expected number of galaxies in the cell.
Moments of the counts in this cell can be  
derived from the moment generating function 
\be
M_{m_{i}}(t)=\int p_{i}(m)\,{\rm e}^{mt} {\rm d}m
=\exp\left[\overline{m_{i}}({\rm e}^{t}-1)\right]
\ee
by differentiating
\be
\overline{m_{i}^{n}}=\left. \frac{{\rm d}^{n}M_{m_{i}}}{{\rm d}t^{n}}\right|_{t=0}
\ee
at zero lag.

We now focus on the smoothed density field
\be
d_{i}=\sum_{j} w_{ij}\frac{m_{j}}{S_{j}},
\ee
where $w_{ij}$ is the value of the effective smoothing kernel
(i.e. taking into account the renormalization due to the ratio method)
between cells $i$ and $j$.

Because  
the $\overline{m_{i}}$ 
are independent variables with respect to the sampling process
the moment generating function for $d_{i}$ is simply
\[
M_{d_{i}}(t)=\prod_{j} M_{m_{j}} \left( \frac {w_{ij}}{S_{j}}
t\right)=
\exp\left[\sum_{j} \overline{m_{j}}\left
({\rm e}^{\frac{tw_{ij}}{S_{j}}}-1\right)\right]\nonumber.
\]

Hence, $d_{i}$ is an unbiased estimate of the smoothed
underlying 
field $\hat{\rho}_{i}=\sum_{j}w_{ij}\rho_{j}$.
An unbiased estimate of the mean density $d$ is therefore 
given by
\be
d=\frac{\sum_{i}g_{i}d_{i}}{\sum_{i}g_{i}}
\ee
for arbitrary weights $g_{i}$. 
We choose $g_{i}=\left[{\rm var}(d_{i})\right]^{-1}$ which 
provides
a minimum variance estimator for $d$. 
To compute
\be
{\rm var}(d_{i})=\left<\overline{d_{i}^{2}}\right>
-\left<\overline{d_{i}}^2\right>
\ee
we first average over sampling realizations (overbar) and then 
with respect to the density field $\rho_{i}$ (angular brackets).

We now want to determine the moments
\be
R_{n}=\left< (\hat{\rho}_{i}-1)^{n}\right> ,
\ee
and, in particular, $R_{2}=\sigma^{2}$.
Introducing the moment generating function
\be
M_{d_{i}-\overline{d_{i}}}(t)=\exp\left[ \sum_{j}\overline{m_{j}}
({\rm e}^{\frac{tw_{ij}}{S_{j}}}-1)-\overline{d_{i}}t\right] ,
\ee
the central moments are given by
\be
\label{Dmoments}
D_{i}^{(n)}=\overline{(d_{i}-\overline{d_{i}})^{n}}=\left. 
\frac{{\rm d}^{n}M_{d_{i}-\overline{d_{i}}}(t)}{{\rm d}t^{n}}\right|_{t=0} .
\ee
Note that due to the rescaling we have $\overline{d_{i}}=\hat{\rho}_{i}$.
We can now relate the moments of the measured field with those of the 
underlying density field by using a binomial expansion
\be
\label{binomial}
(d_{i}-1)^{n}=\sum_{k=0}^{n}  \left( \begin{array}{c}
 n \\ k 
\end{array} \right) (d_{i}-\overline{d_{i}})^{k}(\hat{\rho}_{i}-1)^{n-k} .
\ee
In particular, for $n=2$ we find
$\overline{(d_{i}-1)^{2}}=D_{i}^{(2)}+(\hat{\rho}_{i}-1)^{2}.$

Then  
an unbiased estimate for the variance of the smoothed
underlying density 
field is given by
\be
R_{2}=\frac{\sum_{i}h_{i}\left[ (d_{i}-1)^{2}-\left<
D_{i}^{(2)}\right>\right]}{\sum_{i}h_{i}}
\label{bluesky}
\ee
for arbitrary weights $h_{i}$.
Again, we want to estimate $R_{2}$ with minimum variance 
by choosing the weights as
\be
h_{i}=\frac{1}{{\rm var}\left[(d_{i}-1)^{2}\right]}.
\ee
In order to compute these weights
we make again use of equation (\ref{binomial}):
\bea
\overline{(d_{i}-1)^{4}}& = & D_{i}^{(4)}+4D_{i}^{(3)}
(\hat{\rho}_{i}-1)^{2} \nonumber\\
 & & + \,6 D_{i}^{(2)}(\hat{\rho}_{i}-1)^{2}+(\hat{\rho}_{i}-1)^{4} .
\label{binomial4}
\eea
Here the moments $D_{i}^{(n)}$ are given by equation (\ref{Dmoments}) as
\[
D_{i}^{(2)}=\sum_{j}\rho_{j}\frac{w_{ij}^{2}}{S_{j}},
\mbox{\hspace{1cm}} D_{i}^{(3)}=\sum_{j}\rho_{j}\frac{w_{ij}^{3}}{S_{j}^{2}},
\]
\be
D_{i}^{(4)}=3\left( \sum_{j}\rho_{j}\frac{w_{ij}^{2}}{S_{j}} \right) ^{2} + \sum_{j}\rho_{j}\frac{w_{ij}^{3}}{S_{j}^{2}}.
\ee

We now approximate the sums in these expressions 
by replacing the density $\rho_{j}$ by the averaged (smoothed) 
density $\hat{\rho}_{i}$ in the particular region. We then obtain
\[
D_{i}^{(2)}=\hat{\rho}_{i}Y_{i}^{(2)},
\mbox{\hspace{1cm}}
D_{i}^{(3)}=\hat\rho_{i}Y_{i}^{(3)},
\]
\be
D_{i}^{(4)}=3\left( \hat{\rho}_{i}Y_{i}^{(2)} \right) ^{2} + \hat{\rho}_{i}Y_{i}^{(4)},
\ee
where $Y_{i}^{(n)}$ is given by
\be
Y_{i}^{(n)}=\sum_{j}\frac{w_{ij}^{n}}{S_{j}^{n-1}} .
\ee

Of course, this is only strictly correct for a homogeneous density
field, but in the context of this analysis it can be expected to be a good 
approximation. Note that even if 
the weights $h_i$ are obtained
only approximately, the final estimate of $R_2$ remains unbiased.

We can now take the ensemble average of equation (\ref{binomial4})
and compute the weights $h_i$.
Here a well known problem in counts-in-cells analysis 
arises: the determination of second and higher moments involves 
the whole hierarchy of moments. A minimum variance estimate of $R_2$
requires the knowledge of $R_3$ and $R_4$ which in turn depend on
still higher moments. To close this hierarchy we 
assume that the density field is close to Gaussian, which implies
$R_{3}=0$ and $R_{4}=3R_{2}^{\,2}$. With this assumption the weights 
become 
\bea
h_i^{-1}
&=& Y_{i}^{(4)}+
2\left[ Y_{i}^{(2)} \right]^2
+2R_2^{\,2}\nonumber\\
& &+R_2 \left\{ 3 \left[ Y_{i}^{(2)} \right]^2 +4 Y_{i}^{(3)} +4
Y_{i}^{(2)} \right\}.
\eea
An estimate of $R_2$ can now be obtained by iteratively
solving equation (\ref{bluesky}).

Since the $d_{i}$ are strongly correlated an estimate of the 
statistical error in the determination of $R_{2}$ by directly
computing 
the variance of $R_{2}$ in the above formalism is
unrealistic. 
Instead, we estimate the uncertainty in $R_2$ by using an ensemble of
PSCz-like mock catalogues extracted from a N-body simulation.

\section{Genus amplitude and shot noise}

\label{appshot}

The genus amplitude of a Gaussian random field depends only on 
the behaviour of the two-point correlation function of the 
smoothed density field near the origin. 
We therefore estimate $\hat{\xi}(\vec{r})$ 
under the influence of shot noise, 
allowing the noise level to vary across the survey volume.

We start by considering a discrete field 
$m(\vec{r})=\sum_{i}\delta^{\rm (D)}( \vec{r}-\vec{r}_i)$ of points
arising in a Poisson sampling process of some 
underlying density field $\delta_{\rm t}(\vec{r})$.

Allowing for a radial variation of the expected number density 
$S(r)=\left< m({\bf r})\right>$ of
tracers, 
we define the measured density contrast as
\be
\delta(\vec{r})=\frac{m(\vec{r})}{S(r)}-1 .
\ee
This gives rise to a smoothed density field 
\be
\hat{\delta}(\vec{r})=\int W(\vec{r}-\vec{r}')\delta(\vec{r}') \,{\rm d}\vec{r}
\ee
with an autocorrelation function 
\be
\hat{\xi}(\vec{r})=\left< \hat{\delta}(\vec{r}_{0})
\hat{\delta}(\vec{r}_{0}+\vec{r})\right> .
\label{lab1}
\ee

Performing the
ensemble average in equation (\ref{lab1}) with 
Feldman et al.'s \shortcite{Fe94}
relation
\bea
\left< m(\vec{r'})m(\vec{r}'')\right>
=S(r')S(r'')
\left[ 1+\xi_{\rm t}(\vec{r}'-\vec{r}'')\right]
\nonumber \\ 
+S(r')\,\delta^{\rm (D)}(\vec{r}'-\vec{r}'') ,
\eea
we find 
\bea
\label{xihat}
\hat{\xi}(\vec{r})=\int W(\vec{r}')W(\vec{r}-\vec{r}''-\vec{r}')
\xi_{\rm t}(\vec{r}'')
\,{\rm d}\vec{r}'  \,{\rm d}\vec{r}''
+\nonumber \\
\int W(\vec{r}_{0}-\vec{r}')W(\vec{r}_{0}+\vec{r}- \vec{r}')
\frac {1}{S(r')} \,{\rm d}\vec{r}' ,
\eea
where $\xi_{\rm t}(\vec{r})$ is the autocorrelation function of the  
underlying field.

The second term in equation (\ref{xihat}) describes the 
shot noise contribution which varies across the survey volume. 
Hence the dependence on ${\vec{r}_{0}}$ remains even after the ensemble average. 
Assuming that $S(r)$ varies sufficiently slowly over a smoothing
length 
and choosing a Gaussian smoothing kernel for $W(\vec{r})$ we can 
substitute $S(r')$ by  $S(r_{0})$ in the integrand 
because $W(\vec{r}_{0}-\vec{r}')$ is strongly peaked at $\vec{r}_{0}$. 
We then obtain
\be
\label{xihattot}
\hat{\xi}(\vec{r})=\hat{\xi}_{\rm t}(\vec{r})+\hat{\xi}_{\rm shot}(\vec{r}),
\ee
where
\be
\label{xithat}
\hat{\xi}_{\rm t}(\vec{r})=\frac{1}{(2\pi)^{3/2}\lambda^{3}}\int
{\xi}_{\rm t}(\vec{r}'')\,\exp\left[{-\frac{(\vec{r}-\vec{r}'')^{2}}{2\lambda^{2}}}\right]
{\rm d} \vec{r}''
\ee
is the contribution of the underlying density field and
\be
\label{xishothat}
\hat{\xi}_{\rm
shot}(\vec{r})=\frac{1}{(2\pi)^{3/2}\lambda^{3}S(r_{0})}\,\exp\left({-\frac{\vec{r}^{2}}{2\lambda^{2}}}\right)
\ee
stands for the shot noise contribution.

Equation (\ref{xithat}) reads in Fourier space like
\be
\hat{P}_{\rm{t}}(\vec{k})=P_{\rm{t}}(\vec{k})\,\exp\left({-\frac{\vec{k}^{2}\lambda^{2}}{2}}\right).
\ee
Adopting isotropy for the underlying field, we can also write
\be
\label{xithat2}
\hat{\xi}_{\rm{t}}(r)=\int\hat{P}_{t}(k)\frac{\sin(kr)}{kr}{\rm d}^{3}k.
\ee

We now assume that the density field can approximately be taken to be 
Gaussian. 
In this case the amplitude of the genus curve 
depends only on $\hat{\xi}(0)$ and $\hat{\xi}''(0)$ \cite{Ha86}. 
The previous equations 
give the corresponding contributions from the underlying field as
\be
\sigma_{\rm t}^{2}\equiv\hat{\xi}_{\rm t}(0) = \int\hat{P}_{\rm t}(k)\,{\rm d}^{3}k
\ee
and
\be
\hat{\xi}_{\rm t}''(0)=\int \frac{k^{2}}{3}\hat{P}_{\rm t}(k)\, {\rm
d}^{3}k =-\frac{\left<k^{2}\right>\sigma_{\rm t}^{2}}{3} \, ,
\ee
while the shot noise gives rise to
\be
\sigma_{\rm shot}^{2}\equiv\hat{\xi}_{\rm shot}(0) = \frac {1}{(2\pi)^{3/2}\lambda^{3}S(r_{0})}
\ee
and
\be
\hat{\xi}''_{\rm
shot}(0)=-\frac{1}{(2\pi)^{3/2}\lambda^{5}S(r_{0})}=-\frac{\sigma^{2}_{\rm
shot}}{\lambda^{2}}.
\ee

We can now estimate the amplitude of the genus 
curve under the influence of shot noise. It is given by
\bea 
N&=&\frac{1}{(2\pi)^{2}}\left[ \frac{-\hat{\xi}''(0)}{\hat{\xi}(0)}\right] ^{3/2}  \\
 &=&\frac{1}{(2\pi)^{2}}\left[ \frac{\left< k^{2} \right>}{3} +
\frac{\sigma^{2}_{\rm shot}}{\sigma^{2}_{\rm t}
+\sigma^{2}_{\rm shot}} \left(\frac{1}{\lambda^{2}}
-\frac{\left< k^{2} \right>}{3} \right)\right]^{3/2}.\nonumber
\eea

\end{document}